\documentclass[prd,aps,floats,epsfig,eqsecnum,nofootinbib]{revtex4}
\usepackage{graphicx,amsmath,amssymb,verbatim}
\def\D{{\cal D}}

\newcommand{\Z}{ {\mathbb Z}}

\newcommand{\rif}[1]{(\ref{#1})}

\begin{document}
\title{Magnetic field generation from  non-equilibrium  phase transitions.}
\author{D. Boyanovsky}
\email{boyan@pitt.edu} \affiliation{Department of Physics and
Astronomy, University of Pittsburgh, Pittsburgh, Pennsylvania
15260, USA \\and LPTHE, Universit\'e Pierre et Marie Curie (Paris
VI) et Denis Diderot (Paris VII), Tour 16, 1er. \'etage, 4, Place
Jussieu, 75252 Paris, Cedex 05, France}
\author{H. J. de Vega}
\email{devega@lpthe.jussieu.fr} \affiliation{LPTHE, Universit\'e
Pierre et Marie Curie (Paris VI) et Denis Diderot (Paris VII),
Tour 16, 1er. \'etage, 4, Place Jussieu, 75252 Paris, Cedex 05,
France \\and Department of Physics and Astronomy, University of
Pittsburgh, Pittsburgh, Pennsylvania 15260, USA}
\author{M. Simionato}
\email{mis6@pitt.edu} \affiliation{Department of Physics and
Astronomy, University of Pittsburgh, Pittsburgh, Pennsylvania
15260, USA}
\begin{abstract}
We study the generation of magnetic fields during the stage of
 particle production resulting from spinodal
instabilities during phase transitions out of equilibrium. The
main premise is that long-wavelength instabilities that drive the
phase transition lead to strong non-equilibrium charge and current
fluctuations which generate electromagnetic fields. We present a
formulation based on the non-equilibrium Schwinger-Dyson equations
that leads to an {\bf exact} expression for the spectrum of
electromagnetic fields valid for general theories and cosmological
backgrounds and whose main ingredient is the transverse photon
polarization out of equilibrium. This formulation includes the
dissipative effects of the conductivity in the medium. As a
prelude to cosmology we study magnetogenesis in Minkowski
space-time in a theory of $N$ charged scalar fields to lowest
order in the gauge coupling and to leading order in the large $N$
within two scenarios of cosmological relevance. The
long-wavelength power spectrum for electric and  magnetic fields
at the end of the phase transition is obtained explicitly.
 It follows that  equipartition between
electric and magnetic fields {\bf does not hold} out of
equilibrium.   In the case of a transition from a high temperature
phase, the conductivity of the medium severely hinders the
generation of magnetic fields, however the magnetic fields
generated are correlated on scales of the order of the domain
size, which is much larger than the magnetic diffusion length.
Implications of the results to cosmological phase transitions
driven by spinodal unstabilities are discussed.
\end{abstract}
\date{\today}
\maketitle
\tableofcontents % to be eliminated in the final version
\section{Introduction}
There is compelling astrophysical evidence for the existence of
magnetic fields in the Universe~\cite{Parker}. Recent advances in
observational techniques mainly through Faraday rotation (RM)
complemented with an independent measurement of the electron
column density via pulsar dispersion measurements(DM) or X-ray
emission from clusters\cite{Kronberg:1993vk} indicate that the
strength of these astrophysical magnetic fields is of order of
$\mu\mathrm{G}$ and are coherent on scales up to that of galaxy
clusters
$100~\mathrm{Kpc}-1~\mathrm{Mpc}$\cite{Parker,Kronberg:1993vk}.
These magnetic fields are now deemed to play an important role in
the evolution and dynamics of galaxies but their origin is still
largely unknown\cite{Grasso:2000wj,Dolgov:2001ce,widrev,giovannini1}.

The galactic dynamos  are some of the promising mechanisms to
amplify pre-existing
seeds\cite{Kronberg:1993vk,malyshkin,plasmabook,widrev}. In its
simplest conception a linear or kinetic dynamo transfers energy
from differential rotation to the build-up of a magnetic field
with a typical growth rate determined by the rotation period of a
protogalaxy $\sim \mathrm{Gyr}^{-1}$. There are many alternative
versions of dynamo theory, and some of the most promising require
turbulent flows\cite{Kronberg:1993vk,malyshkin}. Dynamos amplify
seeds but an initial seed must be present. The proposals to
explain the initial seeds can be classified as astrophysical or of
primordial cosmological origin. An important astrophysical
mechanism is the Biermann battery (for a recent discussion
see~\cite{malyshkin}) which relies on gradients in the electron
number density and pressure which are in different directions that
can arise behind hydrodynamic shocks\cite{widrev}. The primordial
cosmological mechanisms purport the generation of seed magnetic
fields during different stages of the Early
Universe\cite{TurnerWidrow}. As originally observed by Turner and
Widrow\cite{TurnerWidrow,widrev} the coupling of the
electromagnetic field to gravity is conformally invariant,
resulting in that cosmological expansion \emph{per se} will not
generate primordial magnetic fields, the coupling of charged
fields to gravity is in general not conformally invariant and can
lead to magnetogenesis through the minimal coupling to
electromagnetic fields. These authors studied a wide range of
possibilities for primordial magnetogenesis with encouraging
results. More recently a thorough study of the generation of
(hyper) magnetic fields during inflation concluded that the
amplitude of the seeds on cosmologically relevant scales is
probably too small\cite{Giovannini:2000dj}. These authors included
dissipative effects in the medium via a kinetic approach that
includes the conductivity.

The amplification of electromagnetic fluctuations during
inflation, by the end of inflation and from inflation to reheating
has been studied by several  authors in the context of different
inflationary models\cite{muchos} (see criticisms in
ref.\cite{giovashapo}). Dolgov~\cite{dolgovanomaly} suggested that
the breaking of conformal invariance through the trace anomaly may
also lead to significant magnetogenesis (see also \cite{joy}) and non minimal
electromagnetic couplings has also been
considered\cite{TurnerWidrow,ratra,Grasso:2000wj}.  Other
proposals suggest that primordial magnetic fields can be generated
much later during the electroweak phase transition\cite{otros} or
in connection with dark energy during the epoch of large scale
structure\cite{Lee:2001hj}. For a pioneering proposal for
magnetogenesis from cosmological first order phase transitions see
ref.\cite{hogan}.

While certainly there is no dearth of proposals an important
aspect that is clear from the present body of work is that a
consistent framework to study the generation of primordial
magnetic fields must include consistently (or semi
phenomenologically) the dissipative effects associated with the
high conductivity of the medium. This aspect was already
highlighted in the seminal work of Turner and
Widrow\cite{TurnerWidrow} and again by Giovannini and
Shaposhnikov\cite{Giovannini:2000dj,giovashapo}.

This is clearly more relevant if the medium is a hot and dense
plasma as prevailed during reheating, preheating and most likely
any phase transition associated with particle physics scales.

\vspace{1mm}

{\bf The goals of this article:} In this article we begin a
program to assess the generation of magnetic fields through the
non-equilibrium processes associated with supercooled (second
order) phase transitions. Here we focus on the process of spinodal
decomposition, namely the process of phase separation, domain
formation and growth resulting from a non-equilibrium phase
transition as the mechanism that produces strong charge
fluctuations which lead to magnetogenesis. Although we focus
primarily on spinodal decomposition, similar arguments will apply
for very weak first order phase transitions with small latent heat
and nucleation barriers, since in this case nucleation will be
almost indistinguishable from spinodal decomposition. We study
models in which there is a charged scalar sector coupled to an
abelian (hypercharge) gauge field \emph{without} breaking the
$U(1)$ symmetry associated with the (hyper) charge. The charged
scalar and abelian gauge fields are supposed to be part of a
larger multiplet of fields pertaining to a larger underlying
theory.

We develop the framework to calculate the power spectrum of
magnetic fields generated out of equilibrium directly from the
underlying quantum field theory. An {\bf exact} expression for the
spectrum of magnetic fields generated through non-equilibrium
processes is obtained directly from the non-equilibrium
Schwinger-Dyson equations for the transverse gauge field
propagator [see eq.(\ref{S.fund})]. This expression is general and
valid for all types of charged matter fields, the main ingredient
being the non-equilibrium transverse polarization for the
(hypercharge) gauge field.

 This framework allows to include
the effects of  a conductivity in the medium and is a
generalization of a previous study of photon production out of
equilibrium\cite{Boyanovsky:1999jh}.

 After establishing a
connection between the dynamics in cosmological space times and
the simplified case of flat Minkowski space time with a model for
the phase transition, we apply these methods to study the power
spectrum of magnetic fields during quenched phase transitions. In
order to study the non-perturbative aspects of spinodal
decomposition, such as the growth of unstable modes below the
phase transition we invoke a model of $N$ charged fields in the
large $N$ limit. Most particle physics models (GUT's, SUSY, etc.)
contain a large number of scalar fields, thus warranting the use
of the large $N$ approach.

The purpose of this study is to extract the important and robust
features that lead to the generation of magnetic fields from these
non-equilibrium processes in a simpler setting and to build
intuition into the main physical aspects.

We postpone the application of these methods to the full
cosmological situation to forthcoming articles.

As mentioned above, many mechanisms for primordial magnetogenesis
have been studied previously~\cite{Grasso:2000wj}-\cite{ratra}.
Our main goal here is to provide {\bf reliable quantitative
estimates} for primordial magnetogenesis during out of equilibrium
phase transitions. In particular we focus on the dynamics of
magnetogenesis in a minimally coupled scalar-gauge theory where
the non-equilibrium dynamics is a consequence of spinodal
instabilities. The formulation presented here is very general, it
includes consistently absorptive effects such as the conductivity
and can be implemented in other scenarios.

In section II we present the main physical picture behind the
mechanism of magnetogenesis studied. In section III we introduce
 the model under consideration and discuss the  issue of
gauge invariance.   Section IV discusses the similarity  between
Friedmann-Robertson-Walker cosmologies and flat Minkowsky space
time that warrants a preliminary study of magnetogenesis within
the simpler setting. As a prelude to cosmology, section V  briefly
reviews the main aspects of non-equilibrium dynamics of spinodal
decomposition in the scalar sector in Minkowsky space time, which
are present in the cosmological setting. In section VI we obtain
the {\bf exact} expression for the power spectrum of magnetic
fields directly from the non-equilibrium Schwinger-Dyson equations
for the transverse photon propagator. We compare the
non-equilibrium result to the more familiar equilibrium setting.

In section VII we implement this framework to study magnetogenesis
in two cosmologically relevant settings that purport to model a
phase transition during inflation and a radiation dominated era.
We discuss in detail how the conductivity of the plasma emerges in
this formulation and obtain the energy density on long-wavelength
magnetic fields extracting features that will remain in the full
cosmological problem.

\section{The physical picture}
Whether a symmetry breaking phase transition occurs in local
thermodynamic equilibrium (LTE) or not is a detailed question of
time scales. Two important time scales must be compared, the time
scale of cooling through the critical temperature
$\tau_{cool}=T(t)/\dot{T}(t)$ and the relaxation time scale of a
fluctuation of typical wavelength $k$, $\tau(k)$. If cooling
through $T_c$  occurs on time scales \emph{slower} than the
relaxation scale, i.e, $\tau_{cool} \gg \tau(k)$ then a
fluctuation of wavevector $k$ adjusts to LTE and for these scales
the phase transition occurs in LTE. On the other hand if
$\tau_{cool} \ll \tau(k)$ fluctuations of wavelength $k^{-1}$
cannot adjust to the conditions of LTE and fall out of
equilibrium, i.e, \emph{freeze out} and for these scales the phase
transition occurs on very short time scales like a quench from a
high temperature phase into a low temperature
phase~\cite{boysinglee}. While short wavelength modes typically
have very short relaxation time scales and remain in LTE during
the phase transition, long-wavelength modes undergo critical
slowing down near $T_c$ and their relaxational dynamics becomes
very slow\cite{critslowdown}. Thus as the temperature nears the
critical with a non-vanishing cooling rate, long-wavelength modes
freeze out falling out of LTE, namely  for these modes  the phase
transition occurs out of equilibrium. The non-equilibrium dynamics
below the critical temperature is described by the process of
\emph{spinodal decomposition}: long-wavelength fluctuations become
unstable and grow initially exponentially in time (in Minkowski
space-time). The field becomes correlated within regions
characterized by a time-dependent correlation length $\xi(t)$ and
the amplitude of the long-wavelength fluctuations becomes
non-perturbatively large since the mean-square root fluctuation of
the field will begin to probe the broken symmetry states with
$\langle \phi^2 \rangle \propto 1/\lambda$ with $\lambda$ the
scalar self-coupling. This is the hallmark of the process of phase
separation, the correlated regions correspond to domains, inside
these domains the field is near one of the vacuum states. A
particular time scale, the nonlinear time $t_{nl} \propto
\ln(1/\lambda)$ , determines the transition from a regime of
linear instabilities to one in which the full non-linearities
become important. This time scale roughly corresponds to when the
mean square root fluctuation of the scalar field samples the
broken symmetry states and the phase transition is almost
complete. At this stage the amplitude of the long-wavelength
fluctuations become of $\mathcal{O}(1/{\lambda})$, the phases
freeze out and the long-wavelength fluctuations become
\emph{classical} but stochastic\cite{nuestros}-\cite{destri}.

Consider now the situation where the scalar field carries an
abelian charge and is coupled to the electromagnetic gauge field.
The strong spinodal fluctuations in the scalar field will induce
fluctuations in the current-current correlation function, and
while the expectation value of the current must vanish by
translational and rotational invariance, the current and charge
correlators will have strong non-equilibrium fluctuations. These
current fluctuations will in turn generate a magnetic field with a
typical wavelength corresponding to the wavelength of the
spinodally unstable modes.

This  is the main premise of this article: the spinodal
instabilities which are the hallmark of a non-equilibrium symmetry
breaking phase transition will lead to strong charge and current
fluctuations of the charged scalar fields which in turn,  lead to
the generation of magnetic fields through the non-equilibrium
evolution.

Thus while the mechanism that drives magnetogenesis is the same as
in most other scenarios, namely, the amplification of charge
fluctuations\cite{Giovannini:2000dj,Lee:2001hj}, in this scenario,
the amplification is the result of the spinodal instabilities of
long-wavelength fluctuations during the phase transition.

The formation and growth  of long-wavelength correlated domains
entails strong non-equilibrium charge and density fluctuations.
These would not be possible if there were long-range interactions,
however in a high temperature plasma the long-range Coulomb
interaction is actually screened. In a high temperature plasma the
Debye screening length is $\xi_{D} \propto
(eT)^{-1}$\cite{kapleb}. Thus long-wavelength
fluctuations will not be hindered by the Coulomb interaction which
is effectively screened over very short distances in a high
temperature plasma.

Primordial magnetic field generation by an electric charge
asymmetry was previously studied by Dolgov and
Silk\cite{dolgovsilk} assuming an early stage in which
electromagnetism was spontaneously broken. In that scenario
long-range forces are screened by the effective mass of the gauge
fields arising from the expectation value of the Higgs field. At
high temperature there is also screening from the plasma, which,
however, was not  taken into account in\cite{dolgovsilk}.

The situation that we study is rather different, the gauge
symmetry is \emph{never} spontaneously broken, but long range
forces are screened on scales given by the Debye screening length
$\xi_D\sim (eT)^{-1}$.

The main ingredients that must be developed in order to understand
the generation of magnetic fields through this non-equilibrium
process are:

\begin{itemize}
\item{A consistent framework to compute the spectrum of generated
magnetic field, namely $\langle \vec{B}({\vec k},t)\cdot
\vec{B}({-\vec k},t)\rangle/V$ with $\vec{B}({\vec k},t)$ the
spatial Fourier transform of the Heisenberg magnetic field
\emph{operator} and $V$ the  volume of the system.} \item{We
anticipate that plasma effects must be included to assess the
generation and eventual decay of the magnetic fields. If there is
a large conductivity in the medium the magnetic field will diffuse
but also its generation will be hindered. This will be a point of
particular importance within the cosmological
setting\cite{Giovannini:2000dj,Grasso:2000wj,TurnerWidrow}. }
\end{itemize}

These ingredients will be analyzed in detail below.

\section{The model }

The cosmological setting, in which we are primarily interested,
corresponds to either a symmetry breaking phase transition during
(or leading to) an inflationary stage or in a radiation dominated
Universe. Such phase transition is in principle different from the
electroweak one\footnote{If the electroweak phase transition is
weakly first order, nucleation will be almost indistinguishable
from spinodal decomposition and the phenomena studied here may be
of relevance.} and presumably occurs at a much higher energy
scale, such as the GUT scale $\sim 10^{15}\mathrm{Gev}$ but is
assumed to be described by a particle physics model that includes
many fields with (hyper)-charge either fermionic or bosonic. We
will not attempt to study a particular gauge theory
phenomenologically motivated by some GUT scenario, but will focus
our study on a generic scalar field model in which the scalar
fields carry an abelian charge, which for simplicity we will take
to be electric charge.

We first focus on the generation of magnetic fields through the
non-equilibrium dynamics of the phase transition  in Minkowski
space-time, and we will argue in a later section that the case of
Friedmann-Robertson-Walker cosmologies in conformal time is very
similar to the Minkowski space-time case. Therefore most of the
technical aspects developed within the framework of Minkowski
space-time will be translated \emph{vis-\`a-vis}  to FRW
cosmologies.

A related study was carried out  in reference
\cite{Boyanovsky:1999jh} where dynamical aspects of screening and
photon production were studied both in the case of non-equilibrium
symmetry breaking as well as of parametric resonance.

While some aspects of the dynamics of magnetic field generation
are related to those studied in~\cite{Boyanovsky:1999jh}, there
are important differences that warrant a detailed treatment. In
particular, the results of photon production out of equilibrium
\emph{could} be used if equipartition held, since both electric
and magnetic fields contribute to photon production. However, as
it will be studied in detail below, out of equilibrium
equipartition is no longer valid and the generation of magnetic
fields must be studied in general, including the effects of the
conductivity.

While details may depend on the specific models, we seek to
extract the robust features of the phenomena which will be
captured by  a model of scalar electrodynamics with $N$ charged
scalar fields. [Within the framework of unified theories, this
abelian gauge field should be interpreted as carrying a
hypercharge quantum number]. Furthermore, the scalar sector
contains also a neutral field which plays either the role of the
inflaton or some generic field that rolls to the minimum of the
potential leading to symmetry breaking, in such a way that
electromagnetism is \emph{not spontaneously broken} to describe
the correct low energy sector with unbroken $U(1)_{EM}$. We will
take the neutral and the $N$ complex (charged) fields to form a
scalar multiplet under an $O(2N+1)$ isospin symmetry. As the
neutral field acquires an expectation value this isospin symmetry
is spontaneously broken to $O(2N)$. Since by construction only the
neutral field acquires a non vanishing expectation value under the
isospin symmetry breaking the photon remains massless (it will
obtain a Debye screening mass from medium effects).

The model under consideration is defined by the following
Lagrangian density, \begin{equation} {\cal
L}=\frac12\partial_\mu\sigma\;
\partial^\mu\sigma+D_\mu\phi^\dagger D^\mu\phi-\frac{1}{4}F^{\mu \nu}F_{\mu
\nu}\nonumber -m^2\left(\frac12\sigma^2+\phi^\dagger\phi\right)
-\frac\lambda{4N} \left(\frac12\sigma^2+\phi^\dagger\phi\right)^2
\label{tlg} \end{equation} \noindent where to simplify the
expressions we have introduced the following notation
$$
\phi^\dagger\phi=\sum_{r=1}^N\phi^\dagger_r\; \phi_r \;,\quad
D_\mu\phi^\dagger D^\mu\phi=\sum_{r=1}^N(\partial_\mu+ieA_\mu)
\phi_r^\dagger\;(\partial^\mu-ieA^\mu)\phi_r\;.
$$
Furthermore, anticipating a non-perturbative treatment of the
non-equilibrium dynamics of the scalar sector in a large $N$
expansion, we have rescaled the quartic coupling in such a way as
to display the contributions in terms of powers of $ 1/N $,
keeping $\lambda$ fixed in the large $N$ limit.

{\bf Gauge invariance:} Gauge invariance must be treated very
carefully in an out of equilibrium situation. In this abelian
theory the constraints arising from gauge invariance can be
accounted for in a systematic and fairly straightforward manner
leading to a \emph{gauge invariant description} as follows (for
details see~\cite{gauginv}). There are here two first class
constraints \begin{equation}\label{constraints} \Pi_0 \approx 0
~~;~~\vec{\nabla}\cdot \vec{E}-\rho \approx 0 \end{equation}
\noindent where $\Pi_0$ is the canonical momentum conjugate to the
time component of the vector field  and $\rho$ is the charge
density. The second one is identified with Gauss' law, the
generator of time independent gauge transformations. At the
quantum level, these constraints annihilate the physical states
simultaneously. In a Schr\"odinger representation of the quantum
field theory, the first constraint implies that the
wave-functionals do not depend on $A_0$. The condition that Gauss'
law annihilates the physical states implies that the
wave-functionals are only functionals of particular gauge
invariant combination of the fields described in detail in
ref.~\cite{gauginv}. In a Hamiltonian description this procedure
is similar to the Coulomb gauge but we haste to emphasize that
this procedure is fully gauge invariant. The instantaneous Coulomb
interaction can be replaced, at the level of the path integral, by
an auxiliary field $A_0$ which is the Lagrange multiplier that
enforces Gauss' law and should \emph{not} be identified or
confused with the temporal component of the vector field. The
final result is that this gauge invariant description is
equivalent to using the following Lagrangian density (for details
see~\cite{gauginv}) \begin{eqnarray}\label{gauginvlag} {\cal L}=
&& \frac12\partial_\mu\sigma\;
\partial^\mu\sigma+\partial_\mu\Phi^\dagger   \;
\partial^\mu\Phi+\frac12\partial_\mu\vec A_T\cdot\partial^\mu \vec
A_T+\frac12 (\nabla A_0)^2
-m^2\left(\frac12\sigma^2+\Phi^\dagger\Phi\right)
-\frac\lambda{4N} \left(\frac12\sigma^2+\Phi^\dagger\Phi\right)^2
\nonumber \\&& -ie\vec A_T\cdot\left(\Phi^\dagger\nabla\Phi-
\nabla\Phi^\dagger\Phi\right)- e^2(\vec
A_T^2-A_0^2)\;\Phi^\dagger\Phi-ie\; A_0\left(
\Phi\dot\Phi^\dagger-\Phi^\dagger\dot\Phi\right) \end{eqnarray}
\noindent where $\Phi$ is a gauge invariant \emph{local} field
which is non-locally related to the original fields, and
$\vec{A}_T$ is the transverse component of the vector field
(${\vec \nabla}\cdot {\vec A}_T=0$) and  $A_0$ is a
non-propagating field as befits a Lagrange multiplier, its
dynamics is  completely determined by that of the  charge
density~\cite{gauginv}.

The main point of this discussion is that the framework to obtain
the power spectrum of the generated magnetic field presented below
is fully \emph{gauge invariant}.

\section{Magnetic fields in Friedmann-Robertson-Walker cosmology}

The action for a charged self-interacting scalar field $\phi$
coupled to an abelian  (charge or hypercharge) gauge field
$\mathcal{A}$ in a general relativistic background geometry is
given by

\begin{equation}\label{lagra} S= \int d^4x \sqrt{-g}\left[g^{\mu\nu}
\mathcal{D}_{\mu}\phi^* \mathcal{D}_\nu \phi -m^2 \phi^*\phi
-\frac\lambda{4N} (\phi^*\phi)^2 -\frac{1}{4} \mathcal{F}_{\mu
\nu}\mathcal{F}_{\alpha \beta}g^{\mu \alpha}g^{\nu \beta}\right]
\end{equation} \noindent  where 
\begin{equation}\label{cova}\mathcal{D}_\mu =
\partial_{\mu}-ie \mathcal{A}_\mu \quad \mbox{and} \quad
\mathcal{F}_{\mu \nu} = \partial_{\mu} \mathcal{A}_{\nu} -
\partial_{\nu}\mathcal{A}_{\mu} - i e [\mathcal{A}_{\mu},
\mathcal{A}_{\nu}] \; . 
\end{equation} 
A Friedmann-Robertson-Walker line element
\begin{equation}\label{FRWds} ds^2= dt^2-a^2(t) \; d{\vec x}^2
\end{equation} \noindent is conformally related to a Minkowski
line element by introducing the  conformal time $\eta$ and scale
factor $C(\eta)$ as \begin{equation} d\eta = \frac{dt}{a(t)}~~;~~
C(\eta) = a(t(\eta)) \end{equation} In terms of these the line
element and metric are given by \begin{equation}\label{conformal}
ds^2 = C^2(\eta) \; (d\eta^2 - d{\vec x}^2)~~;~~g_{\mu \nu}=
C^2(\eta) \; \eta_{\mu \nu} \end{equation} \noindent where
$\eta^{\mu\nu}=\mbox{diag}(1,-1,-1,-1)$ is the Minkowski metric.
Introducing the conformal fields
$$
A_{\mu}(\eta,\vec x)=\mathcal{A}(t(\eta),\vec x) \quad , \quad
\Phi(\eta,\vec x) = C(\eta) \; \phi(t(\eta),\vec x)
$$
and in terms of the conformal time, the action now reads
\begin{equation}\label{confoS} S= \int d\eta \;  d^3x \left[\eta^{\mu
\nu}D_{\mu}\Phi^*D_{\nu}\Phi-M^2(\eta) \; \Phi^*\Phi
-\frac\lambda{4N}(\Phi^*\Phi)^2-\frac{1}{4} F_{\mu \nu} \;
F_{\alpha \beta} \; \eta^{\mu \nu}  \; \eta^{\alpha \beta}\right]
\end{equation} \noindent with
\begin{eqnarray} &&M^2(\eta) = m^2 \;  C^2(\eta)- \frac{C''(\eta)}{C(\eta)}
\label{masconf}\\
&&D_{\mu} = \partial_{\mu}-ie A_{\mu} ~~; ~~ F_{\mu \nu} =
\partial_{\mu }A_{\nu}-\partial_{\nu
}A_{\mu}\label{confof} \end{eqnarray} \noindent and the primes
refer to derivatives with respect to conformal time.  Obviously
the conformal rescaling of the metric and fields turned the action
into that of a charged scalar field interacting with a gauge field
in \emph{flat Minkowski space-time} with  the scalar field
acquiring a time dependent mass term. Here we have omitted the
contribution from the conformal anomaly, which has been studied in
ref.\cite{dolgovanomaly}. In particular, in the absence of the
charged scalar field and  neglecting the contribution from the
conformal anomaly, the equations of motion for the gauge field $A$
are those of a free field in Minkowski space time. This is the
statement that gauge fields are \emph{conformally} coupled to
gravity and no generation of electromagnetic fields can occur from
gravitational expansion alone without coupling to other fields or
breaking the conformal invariance of the gauge sector. The
generation of electromagnetic fields must arise from a coupling to
other fields that are not conformally coupled to gravity (like the
scalar field), or by non minimal electromagnetic couplings that
would break the conformal invariance of the gauge fields.

The conformal electromagnetic fields
$\vec{\mathcal{E}},\vec{\mathcal{B}}$ are related to the physical
 $\vec E, \vec B$ fields by the conformal rescaling
\begin{equation}\label{physicalfields}
\vec{\mathcal{E}}=\frac{\vec{E}}{C^2(\eta)}~~;~~\vec{\mathcal{B}}=
\frac{\vec{B}}{C^2(\eta)} \end{equation} \noindent corresponding
to fields of scaling dimension two.

Since the action (\ref{confoS}) is essentially that of coupled
charged scalar fields with gauge fields in Minkowski space time
(with a time dependent mass term for the scalars), much will be
learned about the cosmological problem by first studying the main
dynamical aspects in Minkowski space-time.

\section{Phase transitions in Minkowski space time: a prelude to
cosmology}

Our ultimate goal is  to study the generation of primordial
magnetic fields during phase transitions either in an inflationary
or radiation dominated eras which are the most likely scenarii for
particle physics phase transitions.

During inflation, the phase transition that we consider
corresponds to the dynamics of small or large field models where
the scalar field has a symmetry breaking or an unbroken potential,
respectively. The rolling of the field corresponds to the
expectation value evolving in time towards the minimum of the
potential but also quantum fluctuations growing through the
spinodal or parametric instabilities~\cite{nuestros,eri95,eri97}.
The initial state is generally described by a gaussian
wave-functional~\cite{eri95,eri97}.

In a radiation dominated cosmology, the initial state is that of
local thermodynamic equilibrium at an initial temperature $T
>>T_c$. The cosmological expansion leads to cooling $T(t)=T/a(t)$
and at some given time the temperature equals the critical and the
phase transition occurs. Using finite temperature field theory in
an expanding background geometry, it is shown that the effective
time dependent mass term includes the time dependent temperature
corrections (for details see for example~\cite{scalingboyhec} and
references therein).

While in a cosmological background the temperature falls off with
the scale factor and the transition from the high temperature
phase to the low temperature phase is driven by the expansion, in
Minkowski space time we must model the transition. As described
above, while short wavelength fluctuations  are in LTE, long
wavelength fluctuations will undergo critical slowing down and for
these the phase transition will occur  as a temperature quench
from the high to the low temperature phase.

A rapid (quenched or supercooled) symmetry breaking phase
transition can be modeled in Minkowski space-time by a
time-dependent mass term which changes sign suddenly from positive
to negative \cite{eri97,boysinglee}. This quench approximation can
be relaxed by allowing a mass term that varies continuously with
time with results that are qualitatively similar\cite{bowick} to
the quench scenario.

We consider two different scenarios that we invoke to model a
quenched phase transition either during an inflationary era or a
post-inflation radiation dominated era:

{\bf i):} A vacuum phase transition, in which the initial state
corresponds to a vacuum state of a free scalar theory with a
positive squared mass term $m^2_i>0$. This corresponds to  an
initial pure Gaussian density matrix describing a free field
theory of scalars with a mass $m_i$~\cite{nuestros,eri95,eri97},
this initial state is evolved in time with the Hamiltonian  with
the symmetry breaking potential.

{\bf ii):} A transition during a radiation dominated era. In this
scenario, before the phase transition ($t > t_0$) the system is in
the unbroken high temperature phase and the scalar fields acquire
a positive thermal mass $m^2_T=(\lambda/24+ e^2/8) T^2$ arising
from scalar tadpole and the gauge contributions to the scalar
self-energy,  whereas after the transition ($t\gg t_0$) the system
is in the low temperature phase with negative mass square
$m^2=-\mu^2$.

Thus we model the time-dependence of the mass term as
\begin{equation}\label{mass.term} m^2(t)=M^2\theta(-t)-\mu^2\theta(t)\;.
\end{equation} \noindent where we have chosen the transition to
occur at $t=0$ and with $M^2>0$ and given by
\begin{equation}\label{mass} M^2=\left\{\begin{array}{l}
m^2_i ~\mathrm{for\;the\;vacuum\;case} \\
\frac{\lambda
T^2}{24}+\frac{e^2T^2}{8}~\mathrm{for\;the\;radiation\;dominated\;case}
\end{array}    \right.
\end{equation} Detailed analytical and numerical studies of the dynamics of
phase transitions in expanding cosmologies reveal that the
qualitative features of phase separation and spinodal
decomposition are well captured by this simple
approximation~\cite{nuestros,eri95,eri97,Boyanovsky:2000hs}. We
postpone to a forthcoming article the full study of primordial
magnetic field generation in a cosmological background.

We emphasize that since the $U(1)$ abelian vector field $A_{\mu}$,
namely the photon, couples to the charged scalar fields and the
$O(2N)$ symmetry of the charged sector is \emph{unbroken}, there
is no Higgs mechanism and the photon does not acquire a mass.

\subsection{Scalar fields dynamics}

For completeness and to highlight the  aspects of the
non-equilibrium dynamics most relevant to the generation of
magnetic fields, we summarize the main features of scalar field
dynamics. For further details the reader is referred
to~\cite{eri97,destri}.

As described above the non-equilibrium evolution of
long-wavelength modes begins with the  spinodal instabilities
which in  Minkowski space-time result in an exponential growth of
the amplitudes for long-wavelength fluctuations. This growth makes
the backreaction important after a certain time eventually
shutting-off the instabilities. That is, the non-linearity
shut-off the spinodal growth of the modes\cite{eri97,destri}.
Therefore a non-perturbative treatment of the dynamics is
required. The large $N$ limit of the scalar sector allows such a
systematic treatment, furthermore it is renormalizable and
maintains the conservation laws\cite{largen}.

The gauge contributions to the self-energy of the scalar field can
be separated into the contributions from hard momentum $k \approx
T$ gauge modes and those from the soft modes with momenta $k \ll
T$. The hard momentum modes will remain in local thermodynamic
equilibrium and their contribution to the self-energy of the
scalar field to lowest order results in a thermal mass $ e
T/\sqrt{8}$\cite{kapleb,boyscalarqed}. This contribution
adds to that from the hard scalar modes $k\approx T$ in  the
scalar tadpole diagram that also leads to a thermal mass
squared $\lambda T^2/24$. Both contributions are accounted for in
the thermal mass $m^2_T$ in equation (\ref{mass.term}).

Furthermore, resumming the contribution from the hard modes (in
equilibrium) to the longitudinal photon propagator (namely the Dyson
series) leads to the screening of the instantaneous Coulomb
interaction with a Debye screening length $\xi_D\sim
(eT)^{-1}$\cite{kapleb}. This results in the screening
of long range forces which would otherwise prevent large charge
fluctuations.

 The
contribution from soft modes of the gauge field to the scalar
self-energy will only include non-equilibrium corrections through
the scalar loops in the photon propagator in the self energy.
Therefore the back-reaction of the non-equilibrium dynamics of the
\emph{gauge field} onto the dynamics of the scalar field will
appear at the two-loop level, namely at order $\alpha^2$.
Therefore we will neglect the non-equilibrium back-reaction of the
gauge field upon the dynamics of the scalar field.

The dynamics in leading order in $N$ already reveals the important
non-equilibrium features of the evolution. Furthermore, we will
neglect the back reaction of the vector field on the dynamics of
the scalar field since these effects are suppressed by at least
one power of the fine structure constant $\alpha$ and are
subleading in the large $N$ limit. The dynamics of the scalar
field in leading order in $N$ is presented in
refs.~\cite{destri,eri97,Boyanovsky:2000hs}.

Since symmetry breaking is chosen along the direction of the
neutral field $\sigma$, we write \begin{equation} \sigma(\vec x,t)
= \sqrt{N}\varphi(t) + \chi(\vec x,t) \quad ; \quad \langle
\chi(\vec x,t) \rangle = 0 \label{expecval} \end{equation}
\noindent where the expectation value is taken in the time evolved
density matrix or initial state. The leading order in the large
$N$ limit is obtained either by introducing an auxiliary
field\cite{largen} and establishing the saddle point or
equivalently  by the factorizations\cite{eri97}
\begin{equation}\label{hartreefac} (\Phi^{\dagger}\Phi)^2 \rightarrow 2 \langle
\Phi^{\dagger}\Phi\rangle \Phi^{\dagger}\Phi \quad , \quad \chi
\Phi^{\dagger}\Phi \rightarrow \chi \langle
\Phi^{\dagger}\Phi\rangle \nonumber \end{equation} The non-linear
terms of the $\sigma$ field lead to contributions of
$\mathcal{O}(1/N)$ in the large $N$ limit, and to leading order
the dynamics is completely determined by the $ N $ complex scalars
$\Phi_r$. This factorization that leads to the leading
contribution in the large $ N $ limit makes the Lagrangian for the
scalar fields quadratic (in the absence of the gauge coupling) at
the expense of a self-consistent condition: thus $\phi$ acquire a
self-consistent time dependent mass.

It is convenient to introduce the mode expansion of the charged
fields \begin{equation} \Phi_r(t,\vec x)= \int \frac{d^3
k}{\sqrt{2 \, (2\pi)^3}} \left[ a_r(\vec k) \; f_k(t)\; e^{i\vec
k\cdot \vec x}+ b_r^\dagger(\vec k)\; f^*_k(t)\; e^{-i\vec k\cdot
\vec x} \right]\; , \label{phidecompo} \end{equation}
\begin{equation} \Phi_r^\dagger(t,\vec x)=\int \frac{d^3
k}{\sqrt{2 \, (2\pi)^3}} \left[ b_r(\vec k)\; f_k(t)\; e^{i\vec
k\cdot \vec x}+ a_r^\dagger(\vec k)\; f^*_k(t)\; e^{-i\vec k\cdot
\vec x} \right] \; , \label{phidaggerdecompo} \end{equation} The
equation of motion for the expectation value $\varphi(t)$ for $t
>0$ [see eq.(\ref{expecval})] is given by\cite{eri97,destri} \begin{equation}
\ddot{\varphi}(t)-\mu^2
\varphi(t)+\frac{\lambda}{2}\varphi^3(t)+\frac{\lambda}{2N}\langle
\Phi^{\dagger} \Phi\rangle \; \varphi(t)=0\; . \label{unscaledeqn}
\end{equation} In leading order in the large $N$ limit, the Heisenberg
equations of motion for the charged fields translate into the
following equations of motion for the mode functions for
$t>0$~\cite{eri97,destri} \begin{equation}\label{unsaledeqnsofmot}
\left[\frac{d^2}{dt^2}+k^2-\mu^2+\frac{\lambda}{2}\varphi^2(t)+
\frac{\lambda}{2N}\langle\Phi^{\dagger}\Phi\rangle\right] \;
f_k(t)=0 \; . \end{equation} The initial conditions for the mode
functions are chosen to describe particles of mass $M$, namely
\begin{equation} f_k(0)= \frac{1}{\sqrt{W_k}} \quad ;
\quad\dot{f}_k(0)= -iW_k~f_k(0)\; , \quad W_k=\sqrt{k^2+M^2}\; ;
\label{iniconds}
\end{equation} \noindent where $M^2$ is given in eq.(\ref{mass}).

For the vacuum case the initial state is annihilated by $a_r(\vec
k);b_r(\vec k)$ whereas for the initial state of local
thermodynamic equilibrium at high temperature \begin{equation}
\langle a_r^\dagger(\vec k)a_s(\vec k)\rangle =\langle
b^\dagger_r(k)b_s(k)\rangle =\delta_{rs} \; n_k~~;~~
n_k=\frac1{e^{\frac{W_k}{T}}-1}\label{BE.scalars} \end{equation}
With this choice of the initial state we find
\begin{equation}\label{backreaction} \frac{\lambda}{2N}\langle\Phi^{\dagger}
\Phi\rangle = \frac{\lambda}{4}\int
\frac{d^3k}{(2\pi)^3}\;|f_k(t)|^2\coth\left[\frac{W_k}{2T}\right]\;.
\end{equation} \noindent with $W_k$ given in eq. (\ref{iniconds}) and the
vacuum case is obtained by setting $T =0$ above.

This expectation value is ultraviolet divergent, it features
quadratic and logarithmic divergences in terms of a momentum
cutoff. The divergences are absorbed in a renormalization of the
mass term $\mu^2 \rightarrow \mu^2_R$ and into a renormalization
of the scalar coupling $\lambda \rightarrow \lambda_R$. These
aspects are not relevant for the discussion here and we refer to
refs.\cite{eri97,destri} for details.

After renormalization, $\frac{\lambda}{2N}\langle \Phi^\dagger
\Phi\rangle$ is subtracted twice, and after rescaling of variables
introduced above it is replaced in the equations of motion by
~\cite{eri97,destri} \begin{equation} \frac{\lambda}{2N}\langle
\Phi^\dagger \Phi\rangle \rightarrow
\lambda_R\Sigma(t)=\frac{\lambda_R}{8\pi^2}\int_0^{\infty} q^2\;
dq \Biggr\{ |f_q(t)|^2 \; \coth\left[\frac{W_k}{2T}\right]
-|f_q(0)|^2
+\frac{\Theta(q-\mu^2_R)}{2q^3}\biggl[-\varphi^2(0)+\varphi(t)^2+
\lambda_R\Sigma(t)\biggr] \Biggr\}. \label{gsigma} \end{equation}
\noindent and the mass and coupling are replaced by their
renormalized counterparts. In order to avoid cluttering of
notation we now drop the subscript $R$ for renormalized
quantities, in what follows $\mu;\lambda$ stand for the
renormalized quantities.

We now consider the case of a quench from an initial state in
which $\varphi(0)=\dot{\varphi}(0)=0$ which from
eq.(\ref{unscaledeqn}) entails $\varphi(t)=0$ as a fixed point of
the dynamics\cite{nuestros,eri97,boysinglee}. This initial case
describes either the vacuum case or the case of a transition from
an initial high temperature phase in local thermodynamic
equilibrium.

After renormalization and in terms of dimensionless quantities,
the non-equilibrium dynamics of the charged scalar fields is
completely determined by the  following equations of
motion~\cite{eri97,destri} \begin{eqnarray}
&&\left[\frac{d^2}{dt^2} - \mu^2+q^2+\lambda\Sigma(t)
\right]f_q(t)=0~~;~~ f_q(0)=\frac{1}{\sqrt{W_q}}~~;~~\dot{f}_q(0)
=-iW_q~f_q(0)\;.  \label{modeeqnew} \\
&&\lambda\Sigma(t)=\frac{\lambda}{8\pi^2}\int_0^{\infty} q^2\; dq
\Biggr\{|f_q(t)|^2 \; \coth\left[\frac{W_q}{2T}\right] -|f_q(0)|^2
+\frac{\Theta(q^2-\mu^2)}{2q^3} \lambda\Sigma(t) \Biggr\}.
\label{gsigmanew} \end{eqnarray} \noindent with $W_q$ given by
equation (\ref{mass}).  The full time evolution determined from
these equations has  been studied in refs.~\cite{eri97,destri}. We
summarize the aspects of the non-equilibrium dynamics that are
relevant for the generation of primordial magnetic fields (for
more details see refs.\cite{nuestros,eri95,eri97}).

In the weak coupling limit $\lambda \ll 1$  the back-reaction
$\lambda\Sigma$ can be neglected as long as $\Sigma < 1/\lambda$.
Neglecting the back-reaction it becomes clear that the mode
functions $f_q(t)$ increase exponentially in the band of
\emph{unstable} wavevectors $q<\mu$ with the result,
\begin{eqnarray} f_q(t)& = &a_q
\exp\left(t\sqrt{\mu^2-q^2}\right)+b_q
\exp\left(-t\sqrt{\mu^2-q^2}\right)\;,\quad q < \mu \;, \\
a_q & = & \frac1{2\sqrt{W_q}}\left(1-\frac{i \;
W_q}{\sqrt{\mu^2-q^2}}\right)~~;~~b_q=a_q^*  \label{alfas} \; .
\end{eqnarray} A feature of the solution (\ref{alfas}) is that when the
exponentially damped solution becomes negligible as compared to
the exponentially growing one, namely for $t > \mu^{-1}$, the
phases of the mode functions $f_q(t)$ {\em freeze}, i.e. become
constant in time and are a slowly varying function of $q$ for long
wavelengths, while the amplitude grows exponentially, namely the
long-wavelength modes behave as \begin{equation}\label{LWmodes}
f_q(t) \approx |a_0| \; e^{i\delta_0}~ e^{\mu t}~e^{-\frac{q^2
t}{2\mu}} \quad \mbox{for} \quad q \ll \mu \; . \end{equation}
This feature of the long-wavelength mode functions will play an
important role in the discussion below.

The exponential growth of modes in the unstable spinodal band
$q<\mu$ make the back reaction term $\lambda\Sigma$  to grow and
eventually  compensate and cancel the $-\mu^2$ in the equations of
motion. This will happen at a {\bf nonlinear} time scale defined
by~\cite{eri97} \begin{equation}\label{spin} \lambda\Sigma(t_{nl})
= \mu^2 \; .
\end{equation} Two important aspects are described by $t_{nl}$: i) at this
time scale the phase transition is almost complete since $\lambda
\Sigma(t_{nl}) = \mu^2$ means that $\lambda \langle \Phi^\dagger
\Phi \rangle /2N = \mu^2 $, namely the mean square root
fluctuations in the scalar field probe the manifold of minima of
the potential.

 ii) At $t \sim t_{nl}$  the amplitude of the field is of order $\mu
/\sqrt{\lambda}$ and the non-linearities become very important.
The back reaction $\lambda \Sigma(t)$ becomes comparable to
$\mu^2$ and the instabilities shut-off.  Thus for $ 0 <t \leq
t_{nl}$ the dynamics is described by the \emph{linear} spinodal
instabilities while for $t>t_{nl}$ a full non-linear treatment of
the evolution is required. However,  detailed analytical and
numerical work both in Minkowski and cosmological space
times~\cite{nuestros,eri95,eri97} have shown that the main
features of the non-equilibrium dynamics can be gleaned from the
intermediate time regime that can be studied within the linear
approximation.

Using the approximation (\ref{LWmodes}) for the mode functions in
the intermediate time regime $t_{nl} > t >>\mu^{-1}$ and for long
wavelengths which give the largest spinodal growth, we find
\begin{equation}\label{sigts} \lambda\Sigma(t_{nl}) \approx  \frac{\mu^2 +
M^2}{32\pi^2 \, M \, \sqrt{\mu}} \;
\frac{\sqrt{\pi}}{t_{nl}^{\frac32}} \; e^{2\mu \, t_{nl}} \left\{
\begin{array}{l}
\lambda~\mathrm{for\;vacuum} \\ \\
\sqrt{96\lambda}~\mathrm{for~high~temperature}
\end{array}  \right.
\end{equation} \noindent where we have used $n_q\approx T/W_q$ and the
long-wavelength approximation $W_q \sim W_0 = \sqrt{\lambda/24}~T$
for the high temperature case. This leads to the following
estimate for the nonlinear time to leading logarithmic order in
the weak coupling~$\lambda$ \begin{equation}\label{nonlint} t_{nl}
\sim \frac{1}{2\mu}\left\{\begin{array}{l}
  \ln\left[\frac{32 (\pi/2)^{3/2}}{
\frac\lambda4\left(\frac\mu M+\frac M\mu\right)} \right]~
  \mathrm{for~vacuum} \\
  \ln\left[\frac{32 (\pi/2)^{3/2}}{\sqrt{6\lambda}\;
\left(\frac\mu M+\frac M\mu\right)} \right]~
  \mathrm{for~high~temperature}
\end{array} \right\}  + {\cal O}\left( \log \log \frac{1}{\lambda}\right)
\end{equation} The amplitude of the long-wavelength modes at the nonlinear
time by the end of the phase transition is approximately
\begin{equation}\label{endPT} |f_q(t_{nl})| \sim \frac{1}{\sqrt{\lambda}}
\end{equation} As we will discuss in detail below this
non-perturbative scale will ultimately determine the strength of
the magnetic fields generated during the phase transition.

During the intermediate time regime the equal time correlation
function is approximately  given by \begin{equation}\label{correl}
\langle \Phi^\dagger_q(t)\Phi_q(t)\rangle \approx |a_q|^2 \;
e^{2\mu\;t}\; e^{-\frac{q^2\;t}{\mu}}\;. \end{equation} \noindent
and its Fourier transform for long wavelengths is of the form
\begin{equation}\label{correlation.lenght} G_>(t,t,\vec x)\approx  |a_0|^2 \;
e^{2\mu\;t}\; e^{-\frac{\vec x^2}{2\xi(t)^2}}\ \end{equation}
\noindent which determines the time dependent correlation length
of the scalar field \begin{equation}\label{corrlength} \xi(t) =
\sqrt{\frac{2t}{\mu}}\;. \end{equation} The detailed analysis of
the dynamics in refs.~\cite{eri97,destri} and the discussion of
the main features presented above can be summarized as follows:

\begin{itemize}
\item{At intermediate times $ 0 < t \leq t_{nl} \simeq \frac{1}{2
\, \mu} \ln\frac{1}{\lambda}$ the mode functions grow
exponentially for modes in the spinodally unstable band $q<\mu$.
The phase of these mode functions \emph{freezes}, namely, becomes
independent of time and slowly varying with momentum.}

\item{ At a time scale determined by the nonlinear time the
back-reaction shuts off the instabilities and the phase transition
is almost complete. This can be understood from the following
argument: the back reaction becomes comparable with the tree-level
term (for $t> 0$) when $\frac{\lambda}{2N}\langle \Phi^\dagger
\Phi \rangle \approx \mu^2$. This relation determines that the
mean square root fluctuation of the scalar field probes the minima
of the tree level potential. }

\item{During the spinodal stage the correlation length of the
scalar field grows in time and is given by equation
(\ref{corrlength}). This is interpreted as the formation of
correlated domains that grow in time, and is the hallmark of the
process of phase separation and ordering. This correlation length
will be important in the analysis of the correlation of magnetic
fields later.  }

\item{The large fluctuations associated with the growth of
spinodally unstable modes of the charged fields will lead to
\emph{current} fluctuations which in turn will lead to the
generation magnetic fields. Thus the most important aspect of the
non-equilibrium dynamics of the charged fields during the phase
transition is that large fluctuations of the charged fields
associated with the spinodal instabilities will lead to the
generation of magnetic fields. Since the modes with longer
wavelength are the most unstable the magnetic field generated
through the process of phase separation will be of long
wavelength. Furthermore, we expect that the magnetic field
generated by these non-equilibrium processes will be correlated on
length scales of the same order as that of the charged field
above.  }

\end{itemize}

\section{Magnetic field spectrum }

Having discussed the mechanism of generation of magnetic fields
through phase separation, the next task is to provide a framework
to calculate the magnetic field generated. Notice that the
\emph{expectation value} of the magnetic field in the
non-equilibrium density matrix must vanish by translational and
rotational invariance, \begin{equation} <\hat B^i(t,\vec
x)>_\rho=0
\end{equation} Hence the relevant quantity to focus on is the
equal time correlation function \begin{equation}\label{corrfuncB}
C(t,\vec x)=<\hat B^i(t,\vec x) \hat B^i(t,\vec 0)>_{\rho}\;,
\end{equation} where we sum on repeated indices. $B(t,\vec x)$
above is a \emph{Heisenberg operator} and the expectation value is
in the initial density matrix. Since the coincidence limit of
correlation functions of operators must be defined carefully, we
define the spectrum of the magnetic field as
\begin{equation}\label{def.spectrum.B} S_B(t,k)=\frac12\lim_{t'\to t}\int d^3x
<\{\hat B^i(t,\vec x), \hat B^i(t',\vec 0)\}>_\rho e^{i\vec k\cdot
\vec x}\;, \end{equation} where $\{\;,\;\}$ denotes the
anticommutator.

In thermal equilibrium the density matrix associated with  the
time-independent hamiltonian of the system takes the form
$\rho=e^{-\beta H}/Z$ which is invariant under time translations.
Hence, the spectrum $S_B(t,k)=S_B(k)$ does not depend on time and
can be computed using the methods of thermal field theory. We will
return to this case below to provide a consistency check of the
formulation out of equilibrium.

In terms of the transverse component of the gauge field
$A^i_T(t,\vec k) = \mathcal{P}^{ij}(\vec k) \; A^j(t,\vec k)$ with
the transverse projection tensor $\mathcal{P}^{ij}(\vec
k)=\delta_{ij}-\frac{k_i k_j}{\vec k{}^2}$ the spectrum $S_B(t,k)$
can be written as (with an implicit sum over indices)
\begin{equation} S_B(t,k)= -\frac{i}{2} \int d^3x\; e^{i\vec
k\cdot \vec x} \; k^2 \; \left.\mathcal{D}^{ii}_H(t,t';\vec
x)\right|_{t=t'} \label{bspectrum} \end{equation} \noindent in
terms of  the \emph{symmetric} correlator of the transverse gauge
field \begin{eqnarray} \mathcal{D}^{ij}_H(t,t';\vec x) & = &
\mathcal{D}^{ij~>}(t,t';\vec x)+\mathcal{D}^{ij~<}(t,t';\vec x)
\nonumber \\
\mathcal{D}^{ij~>}(t,t';\vec x) & = & i< A^i_T(t,\vec x)
A^j_T(t',\vec 0)> ~~; ~~\mathcal{D}^{ij~<}(t,t';\vec x)  =  i<
A^j_T(t',\vec 0) A^i_T(t,\vec x)> \label{symmetcorrA}
\end{eqnarray} Where the correlation functions in
(\ref{symmetcorrA}) are Wightmann functions without time ordering
computed in the initial density matrix. The main reason for
introducing the definitions above is that there is a well
established framework for obtaining these correlation functions in
non-equilibrium quantum field theory as discussed below.

From $S_B(t,k)$ we can extract the magnetic energy density \begin{equation}
\label{magn.energy} \rho_B(t)= \int
\frac{d^3k}{(2\pi)^3}\;S_B(t,k) \; , \end{equation} where the
ultraviolet behavior is understood to be regulated in some gauge
invariant manner, for example with dimensional regularization.

From the magnetic energy density we can define an
\textit{effective magnetic field} $B_{eff}(t)$ such that
\begin{equation}\label{B.eff} \frac12B^2_{eff}(t)\equiv \rho_B(t)
\end{equation} In the non-interacting, thermal equilibrium case,
which will be relevant to compare the energy density in the
generated magnetic field to that in the thermal radiation
background, \begin{equation}\label{thermaleq} S_B^{(0)}(k)=k \;
[1+2n(k)]
\end{equation} \noindent  where $n(k)$ is the Bose-Einstein
distribution function. Subtracting the irrelevant vacuum
contribution, the effective magnetic field is then given by the
Stefan-Boltzmann form \begin{equation}\label{SB} \vec
B_{eff}^2{}^{(0)}=\frac{\pi^2T^4}{15}\;. \end{equation} The phase
transition generates a dynamic effective magnetic field $\Delta
B_{eff}(t)$ through the interaction between the charged fields and
the electromagnetic field. Hence \emph{a priori} we would expect
that $\Delta B_{eff}(t)$ can be obtained systematically in a power
series expansion in the electromagnetic coupling $\alpha$, namely
$\Delta B_{eff}(t)= \alpha \Delta B_{eff}^{(1)}(t)+ \alpha^2
\Delta B_{eff}^{(2)}(t)+\cdots $.

This would be the case were it not for the fact that  the DC
conductivity is important to estimate reliably the amplitude and
the correlations of the generated magnetic field.  As it will be
discussed below in a high temperature plasma the conductivity is
determined by fluctuations of the  charged fields with typical
momenta $p \sim T$. These short wavelength modes  are in LTE while
only low momentum modes of the charged fields will fall out of
equilibrium and undergo spinodal instabilities during the phase
transition. The DC conductivity is \emph{nonperturbative and
non-analytic} in $\alpha$ and to leading logarithmic order it is
approximately given by\cite{baym,yaffe}
\begin{equation}\label{conductivity} \sigma = \frac{\mathcal{C}{N}
T}{\alpha\ln\frac{1}{\alpha{N}}}
\end{equation} \noindent with ${N}$ the number of charged fields
and $\mathcal{C}$ a numerical constant of order
one\cite{baym,yaffe}.

Thus while we cannot provide an expression for  the generated
magnetic field as a power series expansion in $\alpha$ because the
presence of the conductivity prevents such expansion, the strategy
that we pursue here  is to treat the long-wavelength
non-equilibrium current fluctuations in perturbation theory in
$\alpha$ while the short wavelength contribution will be accounted
for in the conductivity. This will be discussed below in detail,
but the important point of this discussion is that the
long-wavelength fluctuations that lead to the generation of the
magnetic field will be treated to lowest order in $\alpha$ and to
leading order in the large $N$ limit.

The reliability of this expansion will be guaranteed if
\begin{equation} \frac{\Delta B_{eff}(t)}{B_{eff}^{(0)}} \ll 1
\end{equation} In cosmology the important quantity is the energy
density in long-wavelength magnetic fields on scales $ L $ equal
to or larger than the galactic scales. Assuming rotational
invariance, we introduce the \emph{energy} density of the magnetic
field generated by the non-equilibrium fluctuations
\begin{equation}\label{rholongwave} \Delta\rho_B(L) = \frac{1}{2\pi^2}
\int^{2\pi/L}_0 k^2~\Delta S_B(k,t) \; dk\;. \end{equation} The
quantity of cosmological relevance is
\begin{equation}\label{ratio} r(L)=
\frac{\Delta\rho_B(L)}{\rho_\gamma} \end{equation} \noindent where
$\rho_\gamma =\pi^2 T^4/15$ is the energy density in the thermal
equilibrium background of photons.

As discussed in section IV, the physical magnetic field in a
cosmological background is diluted by the expansion as $B \sim
1/C^2(\eta)$ with $C(\eta)$ the scale factor. Therefore, in the
absence of  processes that generate or dissipate magnetic fields
 the ratio (\ref{ratio}) would be constant
under cosmological expansion because of the redshift of the
temperature $T \sim 1/C(\eta)$. In a cosmological background the
ratio (\ref{ratio}) will only depend on time through the
generation or dissipative mechanisms (such as magnetic diffusion
in a conducting plasma) but not through the cosmological
expansion.

For $r \geq 10^{-32}$ the linear (kinetic) dynamo may be
sufficient to amplify cosmological seed magnetic fields and for
$r\geq 10^{-8}$ the collapse of protogalaxies with constant
magnetic flux may be sufficient to amplify the seed magnetic
fields\cite{TurnerWidrow,Grasso:2000wj,Dolgov:2001ce,giovannini1}.

In the cosmological setting this ratio is approximately constant
if the conductivity in the plasma is very large after the
non-equilibrium dynamics has taken place. The approximations
invoked to estimate the spectrum of the generated magnetic field
will be reliable provided $r(L) \ll 1$.

\subsection{Non-equilibrium dynamics of electromagnetic
fluctuations}\label{non.eq.methods}

The formulation of quantum field theory out of equilibrium is by
now well established in the literature and we refer the reader to
refs~\cite{ctp,eri95,eri97} for details.  The generating
functional of real-time non-equilibrium correlation functions can
be written as a path integral along a contour in (complex) time.
The forward and backward branches of this contour represent the
time evolution forward and backward as befits the time evolution
of an initial density matrix\cite{ctp,eri95,eri97}. Gathering all
bosonic (scalar and vector) fields generically in a multiplet
$\mathbf{\Psi}$ the generating functional of correlation functions
is given by~\cite{ctp,eri95,eri97}
\begin{equation}\label{genefunc}
Z(J_{+},J_{-})=\int[\mathcal{D}\Psi_{+}\mathcal{D}\Psi_{-}]\; e^{i
\int d^4x \left\{\mathcal{L}(\Psi_{+})-\mathcal{L}(\Psi_{-})+
J_{+}\Psi_{+}- J_{-}\Psi_{-}\right\}}\;. \end{equation} \noindent
where functional derivatives with respect to the sources $J_{\pm}$
lead to the non-equilibrium real time correlation functions. The
doubling of fields with labels $\pm$ is a consequence of the fact
that the non-equilibrium generating functional corresponds to
forward $(+)$ and backward $(-)$ time evolution and suggests
introducing a compact notation in terms of the  doublet
$(\Psi_{+},\Psi_{-})$ and the  metric in internal space~\cite{ctp}
\begin{equation}\label{cab} c^{ab}=\begin{pmatrix}1&0\cr
0&-1\end{pmatrix}=c_{ab}^{-1}\;, \end{equation} \noindent where
the labels $a,b= \pm$. This notation is particularly useful to
obtain the non-equilibrium version of the Schwinger-Dyson
equations for the propagators~\cite{ctp}.

In the case under consideration the main ingredients for the
non-equilibrium description are the following

\begin{figure}[ht!]
\includegraphics[width=3.75in,keepaspectratio=true]{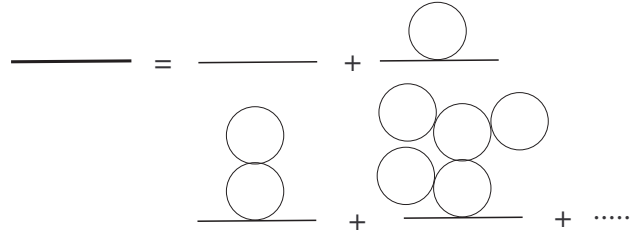}
\caption{ Scalar propagators in leading order in the large N. For
vanishing expectation value of $\sigma$ these are given by a sum
of cactus-type diagrams.} \label{fig:cactus}
\end{figure}

\begin{itemize}
\item{ {\bf Transverse photon propagators}

The real time Green's functions for transverse photons are given
by \begin{equation}\label{photonprops}
{\langle}{A}^{(a)}_{Ti}(\vec{x},t){A}^{(b)}_{Tj}(\vec{x^\prime},
t^{\prime}){\rangle}=-i\int \frac{d^3k}{(2\pi)^3} \;{\mathcal
D}_{ij}^{ab}(k;t,t^\prime)\;e^{-i\vec{k}\cdot(\vec{x}-\vec{x^\prime})}\;,
\end{equation} where the explicit form of ${\mathcal D}_{ij}^{ab}
(k;t,t^\prime)$ is \begin{eqnarray} &&{\mathcal
D}_{ij}^{++}(k;t,t^{\prime})={\cal P}_{ij}(\vec{k}) \;
\left[{\mathcal D}_{ij}^{>}(k;t,t^{\prime})\Theta(t-t^{\prime})
+{\mathcal D}_{ij}^{<}(k;t,t^{\prime})\Theta(t^{\prime}-t) \right]\;, \label{phot++}\\
&&{\mathcal D}_{ij}^{--}(k;t,t^\prime)= {\cal P}_{ij}(\vec{k}) \;
\left[{\mathcal D}_{ij}^{>}(k;t,t^{\prime})\Theta(t^{\prime}-t)
+{\mathcal D}_{ij}^{<}(k;t,t^{\prime})\Theta(t-t^{\prime})
\right]\;,
\label{phot--}\\
&&{\mathcal D}_{ij}^{+-}(k;t,t^\prime)={\cal P}_{ij}(\vec{k}) \;
{\mathcal D}_{ij}^{<}(k;t,t^{\prime})\;; \; {\mathcal
D}_{ij}^{-+}(k;t,t^\prime)={\cal P}_{ij}(\vec{k}) \; {\mathcal
D}_{ij}^{>}(k;t,t^{\prime}) \label{photpm} \end{eqnarray} and
${\cal P}_{ij}(\vec{k})$ is the transverse projection operator,
\begin{equation} {\cal
P}_{ij}(\vec{k})=\delta_{ij}-\frac{k_ik_j}{k^2}\;.\label{projector}
\end{equation} } \item{ {\bf Scalar propagators:}
\begin{equation}\label{scalarprops}
{\langle}{\Phi_r}^{(a)\dagger}(\vec{x},t){\Phi_s}^{(b)}(\vec{x^\prime},
t^{\prime}){\rangle}=-i\delta_{rs}\int \frac{d^3k}{(2\pi)^3}\;
G_k^{ab} (t,t^\prime) \;
e^{-i\vec{k}\cdot(\vec{x}-\vec{x^\prime})}\;, \end{equation}
\begin{eqnarray}
&&G_k^{++}(t,t^\prime)=G_k^{>}(t,t^{\prime})\Theta(t-t^{\prime})
+G_k^{<}(t,t^{\prime})\Theta(t^{\prime}-t)\; , \label{gplpl}
\\
&&G_k^{--}(t,t^\prime)= G_k^{>}(t,t^{\prime})\Theta(t^{\prime}-t)+
G_k^{<}(t,t^{\prime})\Theta(t-t^{\prime})\;, \label{gpll}
\\
&&G_k^{+-}(t,t^\prime)=G_k^{<}(t,t^{\prime})~~; ~~
G_k^{-+}(t,t^\prime)=G_k^{>}(t,t^{\prime}),
\label{gplmin}\\
&&G_k^{>}(t,t^{\prime})=\frac i2\left\{[1+n_k]
f_k(t)f^*_k(t^\prime)+n_k f_k(t^\prime)f^*_k(t)\right\}\;
,\label{greater}
\\
&&G_k^{<}(t,t^{\prime})=\frac i2\left\{\left[1+n_k
\right]f_k(t^\prime)f^*_k(t)+n_k f_k(t)f^*_k(t^\prime)\right\}\;
.\label{lesser} \end{eqnarray} For the scalar propagators we have
used the expansion in terms of mode functions given by eq.
(\ref{phidaggerdecompo}) and the expectation values given by eq.
(\ref{BE.scalars}).  }
\end{itemize}
The scalar propagators given above imply a non-perturbative sum of
cactus-type diagrams when the expectation value of the $\sigma$
field vanishes. These propagators are depicted in  fig.
\ref{fig:cactus}.

\subsection{Non-equilibrium Schwinger-Dyson equations}

In this section we derive an {\bf exact} expression for the
spectrum of the magnetic field from the non-equilibrium set of
Schwinger-Dyson equations.

The full non-equilibrium propagator for the photon field is
obtained from the non-equilibrium effective action resulting from
integrating out the charged fields, and which up to quadratic
order in the photon field is given by
\begin{equation}\label{effectiveaction} \Gamma[A^{\pm}]= \int d^4x
d^4y \left\{A^{a}_{i,T}(x)\left[\partial_{\mu}\partial^\mu
\delta^4(x-y) c^{ab}\delta_{ij}+\Pi^{ab}_{ij,T}(x,y)
\right]A^{b}_{j,T}(y) \right\} \end{equation} \noindent with an
implicit sum over all repeated indices.

\begin{figure}[ht!]
\includegraphics[width=2.75in,keepaspectratio=true]{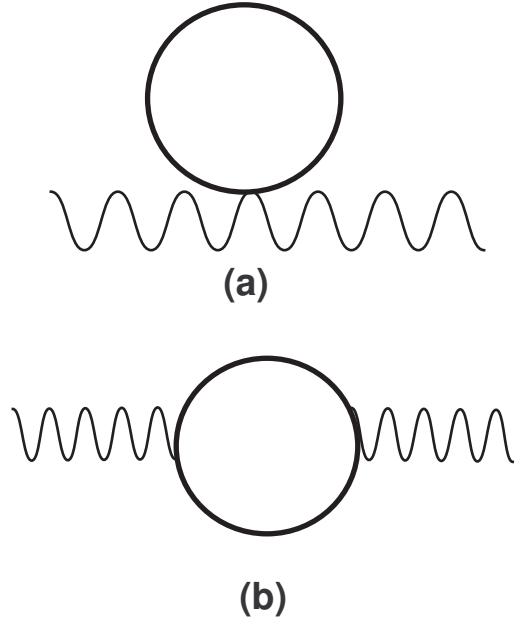}
%\centerline{ \epsfig{file=photonpropagators.eps,width=2in}}
\caption{ Photon polarization to lowest order in $\alpha$ and to
leading order in the large N. The scalar loops are in terms of the
scalar propagators displayed in fig.\ref{fig:cactus}}
\label{fig:polarization}
\end{figure}

To lowest order in $\alpha$ and to leading order in the large $N$
expansion, the non-equilibrium contribution from the scalar fields
to  the photon polarization is given by the two diagrams shown in
fig. \ref{fig:polarization}.

The tadpole diagram (a) in fig.\ref{fig:polarization} gives a
contribution that is \emph{local} in time, thus we write the
polarization \begin{equation}\label{polar}
\Pi^{ab}_{ij,T}(x,y)={\Pi}^{tad}(t) \; c^{ab} \; \delta_{ij} \;
\delta^4(x-y)+ {\mathbf{\Pi}}^{ab}_{ij,T}(\vec x-\vec y,x^0,y^0)
\; , \end{equation} \noindent where the form of the local
(tadpole) contribution ${\Pi}^{tad}(t)$ in terms of the metric
$c^{ab}$ is necessary for consistency with the form of the
effective action since the kinetic term is proportional to the
metric~\cite{ctp} and we have used spatial translational
invariance. The contribution ${\mathbf{\Pi}}$ is non-local, and in
equilibrium it features absorptive parts. The different components
$a,b = \pm$ are defined in the same manner as the Green's
functions (\ref{phot++})-(\ref{photpm}) and
(\ref{gplpl})-(\ref{gplmin}). The contribution to order $\alpha$
to the non-local part of the polarization is depicted in diagram
(b) in fig.\ref{fig:polarization}.

The photon propagator given in eq. (\ref{photonprops}) is the
inverse of the operator in the quadratic effective action above
and obeys the non-equilibrium version of the Schwinger-Dyson
equation~\cite{ctp} \begin{equation}\label{SDeqn} \int d^4y
\left[\partial_{\mu}\partial^\mu \delta^4(x-y) \;  c_{ab} \;
\delta_{ij}+\Pi^{ab}_{ij}(x,y) \right]\mathcal{D}^{jk}_{bc}(y,z) =
\delta^{ik} \; c_{ac} \; \delta^4(x-z) \end{equation} From the
expressions (\ref{phot++})-(\ref{photpm}) for the different
components of the photon propagator it proves convenient to
introduce the following combinations
\begin{eqnarray}\label{definitions}
&&\mathcal{D}^C_{ij}(x,y)=\mathcal{D}^>_{ij}(x,y)-
\mathcal{D}^<_{ij}(y,x)~~;~~\mathrm{antisymmetric~
propagator} \label{antisym} \\
&&\mathcal{D}^R_{ij}(x,y)=\mathcal{D}^C_{ij}(x,y) \;
\theta(x^0-y^0)~~;~~\mathrm{retarded~
propagator} \label{ret}\\
&& \mathcal{D}^A_{ij}(x,y)=\mathcal{D}^C_{ij}(x,y) \;
\theta(y^0-x^0) ~~;~~\mathrm{advanced~
  propagator} \label{adv}\\
&&
\mathcal{D}^H_{ij}(x,y)=\mathcal{D}^>_{ij}(x,y)+\mathcal{D}^<_{ij}(y,x)~~;~~\mathrm{symmetric~
 propagator}\label{sym}.
\end{eqnarray} \noindent and similar definitions for the polarization
$\mathbf{\Pi}$.

Using spatial translational invariance we can simplify the above
form of the Schwinger-Dyson equations by introducing the spatial
Fourier transform of the transverse propagators
\begin{equation}\label{FT} \mathcal{D}^{ab}_{ij,T}(t,t';\vec
x-\vec x')=\int\frac{d^3 k}{(2\pi)^3} \; e^{i\vec k\cdot(\vec
x-\vec x')} \; \mathcal{P}_{ij}(\vec k) \; \D^{ab}(t,t';\vec k)
\end{equation} and introducing the spatial Fourier transform of the
transverse polarization tensors defined by
\begin{equation}\label{FTSE} {\Pi}^{ab}_{T}(t,t';\vec x-\vec
x')=\frac{1}{2}\int\frac{d^3 k}{(2\pi)^3} \; e^{i\vec k\cdot(\vec
x-\vec x')} \; \mathcal{P}_{ij}(\vec k) \; \Pi^{ab}_{ij}(t,t';\vec
k)  \; . \end{equation} We can now write down the Schwinger-Dyson
equations for the transverse photon propagators.  Since there are
only two independent functions $D^{>,<}$ the Schwinger-Dyson
equations (\ref{SDeqn}) can be simplified to a set of two
equations for a particular combinations of these. Some
straightforward algebra leads to the following set of
non-equilibrium Schwinger-Dyson equations (we have suppressed the
subscript $T$ but the equations below refer to the transverse
components) \begin{equation} \label{EQ.DR.k} \int
dt_1\left[\left(\frac{d^2}{dt^2_1}+k^2+{\Pi}^{tad}(t)
\right)~\delta(t-t_1)+\Pi_R(t,t_1;\vec
k)\right]\mathcal{D}_R(t_1,t';k)=\delta(t-t') \end{equation} and
\begin{equation}\label{EQ.DH.k} \int
dt_1\left[\left(\frac{d^2}{dt^2_1}+k^2+{\Pi}^{tad}(t)
\right)~\delta(t-t_1)+\Pi_R(t,t_1;k)\right]
\mathcal{D}_H(t_1,t';k)=\int dt_1\;\mathbf{\Pi}_H(t,t_1;\vec
k)\D_A(t_1,t';k)\;. \end{equation} \noindent with the definition
\begin{equation}\label{retpol} \Pi_R(t,t_1;k)= \left[\Pi^>(t,t_1;\vec
k)-\Pi^<(t,t_1;k) \right]\Theta(t_1-t) \end{equation} A remarkable
aspect of this set of equations is that the right hand side of eq.
(\ref{EQ.DH.k}), namely the inhomogeneity in the set of equations,
only involves the \emph{non-local} contribution to the
polarization, which to lowest order in $\alpha$ is given by
diagram (b) in fig. \ref{fig:polarization}. This is a result of
the form of the local (tadpole) contribution which is proportional
to the metric $c^{ab}$. This is an important point to which we
will come back later in the discussion.

The antisymmetric propagator $ \mathcal{D}_C(t,t'';k) $ is the odd
solution of the homogeneous evolution equation \begin{equation}
\label{ecDC} \left(\frac{d^2}{dt^2}+k^2 +
{\Pi}^{tad}(t)\right)\mathcal{D}_C(t,t'';k) +\int_{t}^{t''}dt' \;
\Pi_R(t,t',k) \; \mathcal{D}_C(t',t'',k)=0  \; , \end{equation}
with the constraint \begin{equation} \label{vincu} \left. \frac{
\partial \mathcal{D}_C(t,t',k)}{\partial t} \right|_{t' = t } = 1
\; . \end{equation} This relation guarantees the correct equal
time canonical commutators.

Since the kernel of the integral equation on the left hand side of
eq. (\ref{EQ.DH.k}) is the same as for the equation that defines
the retarded propagator (\ref{EQ.DR.k}), the solution to eq.
(\ref{EQ.DH.k}) is given by \begin{equation} \label{DH}
\mathcal{D}_H(t,t';k)=\int dt_1\int dt_2\;\mathcal{D}_R(t,t_1;k)
\mathbf{\Pi}_H(t_1,t_2;\vec k)\mathcal{D}_A(t_2,t';k)+F(t,t';k)
\end{equation} \noindent where $\mathcal{D}_R(t,t';k)$ is the solution of eq.
(\ref{EQ.DR.k}) and the function $F(t,t';k)$ symmetric in $t$ and
$t'$ is a general solution of the homogeneous equation
\begin{equation}\label{H.E.} \int
dt_1\left[\left(\frac{d^2}{dt^2_1}+k^2+{\Pi}^{tad}(t)
\right)~\delta(t-t_1)+\Pi_R(t,t_1;k)\right]F(t_1,t',k)=0\;.
\end{equation} The homogeneous solution $F$ can be constructed
systematically and its physical significance will be discussed
below.

Now are now in position to provide the final expression for the
 spectrum of the magnetic field. From the expression
 (\ref{bspectrum}) and the solution found above, we find
\begin{equation} S_B(t,k)= -ik^2\left.\mathcal{D}^H(t,t';k)\right|_{t=t'}
\label{bspectrumfin} \end{equation} The expression (\ref{DH}) can
be simplified further by taking into account the theta  functions
in the definitions of the retarded and the advanced ($\D_R$ and
$\D_A$) propagators [see eqs. (\ref{ret})-(\ref{adv})] as well as
the antisymmetry $\D_C(t,t';\vec k)=-\D_C(t',t;k)$ of
$\mathcal{D}_C$ leading to \begin{equation} \label{finalBcor}
S_{B}(t,k)=i~k^2\left\{\int_{t_0}^t dt_1\int_{t_0}^{t}
dt_2\;\D_C(t,t_2;k)\; \D_C(t,t_1;k) \;
\mathbf{\Pi}_H(t_1,t_2;k)-F(t,t;k)\right\}\;. \end{equation} Since
the product $\D_C(t,t_2;k)\D_C(t,t_1;k)$ is symmetric in the
exchange $t_1\leftrightarrow t_2$ we can replace
$\mathbf{\Pi}_H(t_1,t_2;k)$ by $2\mathbf{\Pi}_>(t_1,t_2;\vec k)$
and write the final form of the spectrum separating, for further
convenience, the contribution from the inhomogeneous and
homogeneous solutions to $S_B$. \begin{eqnarray}\label{S.fund}
&& S_B(t,k) = S^I_B(t,k)+ S^H_B(t,k)\nonumber \\
&& S^I_B(t, k)=2i~k^2~\int_{t_0}^t dt_1\int_{t_0}^{t}
dt_2\;\D_C(t,t_2;k)\; \D_C(t,t_1;k)\;
\mathbf{\Pi}_>(t_1,t_2;k)\nonumber \\
&&S^H_B(t,k)=-ik^2~F(t,t;k)\;. \end{eqnarray} \noindent where
$t_0$ is  some initial time before the phase transition and
$\mathbf{\Pi}_>$ \emph{does not} include the local tadpole
contributions, it is the \emph{non-local} part of the
polarization.

There is an important aspect associated with the homogeneous
solution $F(t,t',k)$ and its contribution to the spectrum of
generated magnetic fields $S^B_H(t,k)$. This aspect is revealed by
noticing that the \emph{expectation value} of the transverse gauge
field $\mathcal{A}_T(t,k)= \langle A_T(t,\vec k)\rangle $, namely
the \emph{mean field} obeys the same homogeneous equation of
motion as $F(t,t',k)$, \begin{equation}\label{gaugeeqn} \int
dt_1\left[\left(\frac{d^2}{dt^2_1}+k^2+{\Pi}^{tad}(t)
\right)~\delta(t-t_1)+\Pi_R(t,t_1;\vec
k)\right]\mathcal{A}_T(t,k)=0 \end{equation} \noindent (we have
suppressed the vector indices to avoid cluttering of notation).

Thus the homogeneous solution $F(t,t',k)$ can be constructed out
of the independent solutions of the mean field equations of motion
(\ref{gaugeeqn}). The main reason that we bring up this point is
to highlight that the solutions to the mean-field equations of
motion are only \emph{part} of the contributions to the generation
of magnetic fields through non-equilibrium processes. However, as
it will be discussed in detail in the next section, this
contribution can be \emph{neglected} in the present case of
non-equilibrium spinodal decomposition in many circumstances, and
 the term $S^I_B(t,k)$ dominates for late times.

\subsection{Electric fields:}

For completeness we now address the generation of electric fields.
The importance of generation of electric fields is mainly related
to the question of equipartition. It is often \emph{assumed} that
the energy density stored in electromagnetic fields is equally
partitioned between electric and magnetic field components, namely
between temporal and spatial gradients. While this is usually the
situation in \emph{equilibrium}, it is not necessarily the case
strongly out of equilibrium such as the situations envisaged in
this article.

The electric field is  the Hamiltonian conjugate field to the
vector potential, and its transverse component is given by
$E_T^i(t,\vec k)=-\dot A_T^i(t,\vec k)$, therefore the equal time
correlation function of the electric field is given by
\begin{equation} \label{efield} S_{E_T}(t,\vec x)=<\dot
A_T^i(t,\vec x)\dot A_T^i(t,\vec 0>_c=
\partial_t\partial_{t'}\left.<\frac12\{A_T^i(t,\vec x),A_T^i(t',\vec
0)\}>_c\right|_{t'=t}\;. \end{equation} \noindent where just as in
the case of the magnetic field we wrote the equal time correlator
as the symmetrized connected two-point correlation function.

Following the steps leading to eq. (\ref{finalBcor}) for the
magnetic field, and using the following identities
$$
\partial_t \D_R(t,t_1;k)=\partial_t \D_C(t,t_1;k) \; \theta(t-t_1),\quad
\partial_{t'} \D_A(t_2,t';k)=-\partial_{t'} \D_C(t',t_2;k) \;
\theta(t_2-t')
$$
we now find \begin{equation}\label{finalEcor} S_{E_T}(t,
k)=\partial_t\partial_{t'}\left\{2i\int_{t_0}^t dt_1\int_{t_0}^{t}
dt_2\;D_C(t,t_2;k)D_C(t',t_1;k) \Pi_>(t_1,t_2;k)- F(t,t',k)
\right\}_{t=t'} \end{equation} The number of photons produced
through the non-equilibrium processes is given by
\begin{equation}\label{def.N}
N(t,k)=\frac{S_{E_T}(t,k)+S_B(t,k)}{2k}-1\;, \end{equation}
\noindent after summing over the two polarization states.

The final form for the spectrum of magnetic and electric fields
given by eq. (\ref{S.fund})-(\ref{finalEcor}) as well as the
number of photons given by eq. (\ref{def.N}) are the main tool to
compute the spectrum of electromagnetic fields generated during
non-equilibrium processes and one of the main results of this
article.

We emphasize that these expressions are {\bf exact} and general
and apply (with rather minor modifications as explained in a
previous section) in the cosmological setting. They allow to study
the problem of the generation of magnetic fields through
non-equilibrium processes in general situations.

Furthermore, the final expressions for the spectrum of
electromagnetic fluctuations (\ref{S.fund}, \ref{finalEcor}) is
valid more generally in spinor electrodynamics since it only
involves the full polarizations and the Schwinger-Dyson equations
for the correlation functions which ultimately lead to the final
expressions are general.

\subsection{Spectrum of fluctuations in equilibrium}

Before focusing on the study of the magnetic field generated
during non-equilibrium phase transitions, it is both illuminating
as well as important as a consistency check to address the case of
thermal equilibrium. In this case the  simplest manner to compute
the spectrum is by using the imaginary time or Matsubara
formulation. where the transverse photon propagator is written
as~\cite{kapleb} 
\begin{equation} 
\D^{ij}_E(\tau,\vec
x)=<A^i_T(\tau,\vec x)A^j_T(\tau',\vec 0)>=\int
\frac{d^3k}{(2\pi)^3} \;  e^{i{\vec k}\cdot{\vec x}} \; T \; \sum_{n\in
\Z}\mathcal{P}^{ij}(\vec k) \; \D_T(\omega_n,k) \; e^{-i\omega_n
(\tau-\tau')}~~;~~\omega_n=2\pi T~n \end{equation} \noindent with
\begin{equation} \D_T(\omega_n,k) = \int dk_0 \;
\frac{\rho_T(k,k_0)}{k_0-i\omega_n} 
\end{equation} 
\noindent and
$\rho_T(k,k_0)$ is the spectral density for transverse photons
which is an odd function of $k_0$. The spectrum for the magnetic
field is obtained from the equal time limit of the Euclidean
propagator, and is therefore given by
\begin{equation}\label{S.ITF} 
S_B(k)=2 \; k^2 \;  T\sum_{n\in \Z}\D_T(\omega_n,k)\;. 
\end{equation} 
The sum over the Matsubara
frequencies can be performed using the methods described
in~\cite{kapleb} leading to 
\begin{equation}
\label{S.B.eq} S_B(k)=2\int d\omega\; k^2 \; n(\omega) \;
\rho_T(\omega,k)\;. \end{equation} The general form of the
spectral density in terms of the transverse polarization is given
by~\cite{kapleb} 
\begin{equation} \label{rhoT}
\rho_T(k,k_0) = \frac{1}{\pi}
\frac{\mathrm{Im}\Pi_T(k,k_0)}{\left[k^2_0-k^2-\mathrm{Re}\Pi_T(k,k_0)
\right]^2+\left[\mathrm{Im}\Pi_T(k,k_0) \right]^2} 
\end{equation}
The spatial and temporal Fourier transform of the retarded and
advanced transverse photon propagators are given
by~\cite{kapleb} 
\begin{eqnarray} &&
\mathcal{D}^R_{ij}(k,\omega) =\mathcal{P}_{ij}(k)
\mathcal{D}^R(k,\omega)~~;~~\mathcal{D}^R(k,\omega)=
\frac{1}{-\omega^2+k^2+\mathrm{Re}\Pi_T(k,\omega)+ i \;
\mathrm{Im}\Pi_T(k,\omega)} \label{FTret}\cr \cr &&
\mathcal{D}^A_{ij}(k,\omega) = \mathcal{P}_{ij}(\vec
k)\mathcal{D}^A(k,\omega)~~;~~\mathcal{D}^A(k,\omega)=
\frac{1}{-\omega^2+k^2+\mathrm{Re}\Pi_T(k,\omega)-i \;
\mathrm{Im}\Pi_T(k,\omega)} 
\end{eqnarray} 
We are now in
conditions to establish contact with the non-equilibrium result of
the first section. In equilibrium the polarization and the
propagators are functions of the time differences and we can then
take the Fourier transform in time of $\mathcal{D}_H(t,t',k)$
given by eq. (\ref{DH})
\begin{equation}\label{FTDH} \mathcal{D}_H(k,\omega)=
\mathcal{D}_R(k,\omega)\left[\mathbf{\Pi}_>(k,\omega)
+\mathbf{\Pi}_<(k,\omega)\right]\mathcal{D}_A(k,\omega)+
F(k,\omega) \end{equation} \noindent where we have explicitly
written $\mathbf{\Pi}$ in terms of $\mathbf{\Pi}_{<,>}$. In
equilibrium the detailed balance (or KMS) condition relates these
components of the polarization to the imaginary
part\cite{kapleb}
\begin{equation}\label{KMS} 
\mathbf{\Pi}_>(k,\omega)= \frac{i}{\pi}
[1+n(\omega)]  \;
\mathrm{Im}\mathbf{\Pi}(k,\omega)~~;~~\mathbf{\Pi}_<(k,\omega)=
\frac{i}{\pi} \;  n(\omega) \;  \mathrm{Im}\mathbf{\Pi}(k,\omega)
\end{equation} 
\noindent with $n(\omega)$ being the Bose-Einstein distribution function.

Furthermore, the Fourier transform of the homogeneous solution
$F(t,t',k)$ obeys  the Fourier transform of eq. (\ref{H.E.})
namely \begin{equation} \label{FTHOMO} \D_R^{-1}(\omega,k) \;
F(\omega,k)= 0 \end{equation} Combining all the above ingredients
and using the fact that $\mathrm{Im}\mathbf{\Pi}(k,\omega)$ is an
\emph{odd} function of $\omega$\cite{kapleb} we are led
to the conclusion that eq. (\ref{bspectrumfin}) for the spectrum
becomes
\begin{equation} \label{SBk}
S_B(k)=\int d\omega\; \left\{2 \; k^2 \; n(\omega) \;
\rho_T(\omega,k) + k^2\; F(\omega,k)\right\} 
\end{equation}
\noindent with $\rho_T(\omega,k)$ given by eq. (\ref{rhoT}).

This result differs from  that obtained via the equilibrium
propagator eq. (\ref{S.B.eq}) by  the contribution of the
homogeneous solution $F$.

In order to understand the source of the discrepancy  between the
two formulations, namely the contribution of the homogeneous
solution $F(k,\omega)$, let us focus on the defining equation for
$F(k,\omega)$ (\ref{FTHOMO}).

In order for a non-vanishing solution to this equation to exist,
from the expression for $\mathcal{D}_R(k,\omega)$ we infer that
\begin{eqnarray}
&&\omega^2-k^2-\mathrm{Re}\Pi_T(k,\omega)=0 \label{disper}\\
&&\mathrm{Im}\Pi_T(k,\omega)=0 \label{zerowidth} \end{eqnarray}
Eq. (\ref{disper}) determines the \emph{dispersion relation} of
quasiparticles and eq. (\ref{zerowidth}) determines that these
quasiparticles must have zero width, i.e, the solutions of the
homogeneous equations are \emph{propagating} quasiparticles with
zero width and a dispersion relation given by (\ref{disper}). The
homogeneous solution is therefore \begin{equation} F(k,\omega)
\propto \delta(\omega^2-k^2-\mathrm{Re}\Pi_T(k,\omega))
\end{equation} In a non-perturbative resummation of the Dyson
series for the propagators and the spectral density,  the limit
when $\mathrm{Im}\Pi_T \rightarrow 0^\pm$ leads to
\begin{equation} \rho_T \rightarrow
\mp\delta(\omega^2-k^2-\mathrm{Re}\Pi_T(k,\omega)) \end{equation}
\noindent and we recognize in this case that the possible
contributions of the homogeneous solutions are  accounted for in
the expressions (\ref{rhoT}) and arise from the propagating
states, and to zeroth order account for the contribution to the
spectrum of the magnetic field from \emph{free} photons.

However, in a perturbative expansion perturbation theory is only
reliable \emph{away} from the single particle poles in the
propagator and the limit $\mathrm{Im}\Pi_T \rightarrow 0^\pm$ will
miss the contribution from the isolated poles. The homogeneous
solution $F(k,\omega)$ gives the required contribution, thus
guaranteeing the consistency of the perturbative expansion. Thus
while the homogeneous solution $F$ \emph{must} be included in a
perturbative calculation of the spectrum, it will be accounted for
in a non-perturbative computation that includes a Schwinger-Dyson
resummation for the full propagator and should \emph{not} be
included in the computation of the spectrum.

In a plasma the true degrees of freedom are \emph{quasiparticles},
in particular for the power spectrum of  magnetic and transverse
electric fields, the important degrees of freedom are transverse
\emph{plasmons}. In the hard thermal loop
approximation\cite{kapleb} the transverse plasmons
contribute to the spectral density $\rho_T(k,\omega)$ a pole term
of the form $Z_T(k)[\delta(\omega-\omega_p(k))-
\delta(\omega+\omega_p(k)]$ with $\omega_p(k)$ being the plasmon
dispersion relation\cite{kapleb} with $\omega_p(0)=
eT/3$. For $k \gg eT, \; \omega_p(k)\sim k~;~Z_T\sim 1/2k$ and we
recover the power spectrum of free field theory. However for $k
\lesssim eT$ the power spectrum of the magnetic field reveals the
presence of collective degrees of freedom, $S_B(k) \sim
kZ_T(k)[1+2 \, n(\omega_p(k)]$ for $e^2T \leq k \leq eT$. 

For ultrasoft frequency and momenta $k,\omega \ll eT$ which is the case
of relevance for the cosmological case, the effective low energy
form of the gauge field (transverse) propagator (well below the
plasmon pole) is
\begin{equation}
\D(k,\omega) \simeq \frac{1}{\omega^2-k^2-i\sigma \omega}
\end{equation}  
leading to the small frequency, long-wavelength form of the spectral density
\begin{equation}
\rho_T(k,\omega) \simeq \frac{1}{\pi} \frac{\sigma \; 
\omega}{(\omega^2-k^2)^2+(\sigma \omega)^2} \; .
\end{equation}
Therefore, for $\omega \ll \sigma $
\begin{equation}
k^2 \; n(\omega) \; \rho_T(k,\omega) \simeq \frac{T}{\pi}
\frac{\Gamma}{\omega^2+\Gamma^2} ~~;~~ \Gamma \equiv
\frac{k^2}{\sigma} \; .
\end{equation}
\noindent Thus for long wavelengths $k\ll \sigma \approx T/\alpha$
the spectral density features a pole at zero frequency leading to
a long-wavelength power spectrum
\begin{equation}\label{SBT}
S_B(k\ll \sigma) = T \; ,
\end{equation}
coming from the first term in eq.(\ref{SBk}).

The plasmon pole contributes to $ S_B(k) $ through the homogeneous
solution $F$ in  eq.(\ref{SBk}) yielding,
\begin{equation}\label{SBp}
S_{B plasmon} \simeq
\frac{k^2}{\omega_p(0)}[1+2n(\omega_p(0)]\sim \frac{2 k^2
T}{\omega^2_p(0)} \sim \frac{k^2}{e^2T} 
\end{equation} 
where we have used the long-wavelength limit of the dispersion relation
$\omega_p(k)$ and residue $Z_T(k)$\cite{kapleb}. The contribution
(\ref{SBp}) is clearly much smaller than (\ref{SBT}) 
in the long-wavelength limit $k\ll eT$. Hence, the magnetic
field power spectrum in equilibrium is $S_B(k) \simeq T$.

Having clarified the role of the homogeneous solution and the form
of the power spectrum within the more familiar equilibrium
setting, we return to the non-equilibrium situation.

\section{Spectrum  out of equilibrium}

The set of equations for the spectra of magnetic and electric
fields (\ref{S.fund})-(\ref{finalEcor}) is {\bf exact} and only
involves the transverse polarization from which the full
propagator and  inhomogeneous and homogeneous solutions are
obtained. Obviously in order to make progress and obtain an
estimate for magnetic and electric field generation one has to
make approximations. In what follows we will obtain the spectra
for magnetic and electric fields to lowest order in $\alpha$ and
to leading order in large $N$.

To this order the tadpole (local) and bubble (non-local)
contribution to the polarization are given respectively by
diagrams (a) and (b) in fig. \ref{fig:polarization}. Their
explicit expressions are given by \begin{eqnarray}\label{leadtad}
\Pi^{tad}(t) &=& -ie^2 N \int \frac{d^3q}{(2\pi)^3} \;
{G}_>(t,t,q) \; , \cr \cr \Pi_{ij,>}(t_1,t_2,\vec k)&=&ie^2 N \int
\frac{d^3q}{(2\pi)^3} \; (2q_i+k_i) \; (2q_j+k_j) \;
G_>(t_1,t_2,q) \; G_>(t_1,t_2,|\vec q+\vec k|)\;. \end{eqnarray}
\noindent which leads to the transverse components
\begin{eqnarray}\label{PITranv}
 && \Pi_>(t_1,t_2, k) =2ie^2 N\int
\frac{d^3q}{(2\pi)^3} \; q^2 \; (1-\cos^2 \theta)~G_>(t_1,t_2,q)
\; G_>(t_1,t_2,|\vec q+\vec k|)\; , \cr \cr &&  \Pi_<(t_1,t_2,
k)=\Pi_>(t_2,t_1, k)\; . \end{eqnarray} where $ \cos\theta\equiv
{\hat q}\cdot {\hat k} $. The scalar propagator $G_>(t,t')$  in
terms of the mode functions $f_q(t)$ that satisfy the evolution
equations (\ref{unsaledeqnsofmot}) is given by [see eq.
(\ref{greater})],
\begin{equation}\label{scalpropa} G_>(t_1,t_2;k)=\frac
i2\left[(1+n_k)f_k(t_1)f_k^*(t_2)+n_k f_k^*(t_1)f_k(t_2)\right]\;.
\end{equation} \noindent and we assumed that before the phase transition the
scalar fields have occupation $n_k$ (obviously this assumption can
be relaxed straightforwardly).

From these expressions we can obtain a reliable estimate of the
non-equilibrium effects in the polarization. For intermediate
times after the phase transition, when the dynamics is dominated
by the spinodal instabilities and before the non-linearities in
the scalar field evolution are important, the long-wavelength mode
functions are approximately given by eq.(\ref{LWmodes}).

Near the end of the phase transition for $t \sim t_{nl} \approx
\frac{1}{2 \, \mu} \ln\left(\frac{1}{\lambda}\right)$ the leading
order time dependence of the scalar Green's functions is then
approximately given by \begin{equation}\label{GFts}
G_>(t,t',k)_{t,t'\sim t_{nl}} \approx e^{\mu(t+t')} \approx
\frac{1}{\lambda} \end{equation} \noindent which allows us to
estimate the order of magnitude of the different terms in the
polarization. For the tadpole contribution (local term) we find
\begin{equation}\label{tadts} \Pi^{tad}(t)_{t\sim t_{nl}} \approx
e^2 ~\mu^2 \; e^{2\mu t}\approx e^2  \; \frac{\mu^2}{\lambda} +
{\cal O}\left(\frac{1}{\sqrt{\lambda}}\right) \end{equation} This
estimate is consistent with the fact that the tadpole contribution
is $ e^2 <\Phi \Phi> $ and near the end of the phase transition
the mean square root fluctuations of the scalar field probe the
vacuum state, namely $<\Phi^2> \sim \mu^2 /\lambda$. Similarly, we
find for $k \ll \mu$, \begin{equation}\label{Pigr}
\Pi_>(t,t',k)_{t\sim t_{nl}} \buildrel{t\sim t_{nl}}\over=
-\frac{i \; e^2 N \sqrt{\mu}}{128 \; \pi^{\frac32}}\;  \left( 1 +
\frac{M^2}{\mu^2} \right)^2 \; \frac{e^{(t+t')(2\mu -
\frac{k^2}{4\mu})} }{(t+t')^{\frac52}}\left[1+{\cal
O}\left(\frac{1}{t}\right) \right] \; , \end{equation} \noindent
which for $t\sim t' \sim t_{nl}$ is of the order $1/\lambda^2$.

Since the retarded polarization given by eq. (\ref{retpol}) is
antisymmetric in the time arguments the leading contribution given
by (\ref{Pigr}) above, actually {\bf cancels} since
$\Pi_<(t,t',k)=\Pi_>(t',t,k)$. A detailed analysis (see
ref.~\cite{Boyanovsky:1999jh}) reveals that the retarded non-local
part $\Pi_R$ is of the same order as the tadpole contribution,
namely \begin{equation}\label{pireta} \Pi_R \approx e^2
~\frac{\mu^2}{\lambda} \end{equation} \noindent  and that in fact
at long times there is an exact cancellation between the two terms
for $k \approx 0$ \emph{in Minkowski space-time}. This is in
agreement with the equilibrium result of a vanishing magnetic
mass\cite{kapleb}.

\subsection{Spinodal decomposition from a vacuum state}

To understand the order of magnitude of the generated magnetic
fields and to gain insight into the main features of the
non-equilibrium processes, we now study the situation of spinodal
decomposition from an initial \emph{vacuum state}. This
corresponds to considering an initial state in which the $\sigma$
field rolls down the potential hill while the fluctuations of the
charged fields are in the vacuum state at the initial time. The
dynamics of the scalar field in this case has been studied
thoroughly analytically and numerically in
references~\cite{boysinglee,eri95,eri97,destri} to which the
reader is referred.

For times smaller than or of the order of the nonlinear time
(\ref{nonlint}) we can evaluate the tadpole diagram
(\ref{leadtad}) using eqs.(\ref{scalpropa}) and (\ref{LWmodes})
with the result, \begin{equation}\label{evatad} \Pi^{tad}(t) =
\frac{e^2 \, N}{64 \, \pi^{\frac32}} \; (\mu^2 + M^2)\;
\sqrt{\frac{M}{\mu}} \; (1 + 2 \, n_0 ) \;  \frac{e^{2\, \mu \,
t}}{(M\, t)^{\frac32}} \left[ 1 + {\cal O}\left( \frac{1}{t}
\right) \right]\; . \end{equation} Let us now consider the
retarded bubble diagram \begin{equation} \label{piret}
\Pi_R(t_1,t_2; k) =-4e^2 N\int \frac{d^3q}{(2\pi)^3} \;  q^2 \;
(1-\cos^2 \theta)~ \mbox{Im}\left[ G_>(t_1,t_2,q)\;
G_>(t_1,t_2,|\vec q+\vec k|)\right] \; \theta(t_2-t_1)
\end{equation} where we used eq.(\ref{retpol}) and
(\ref{PITranv}).

We are interested in the generation of long-wavelength magnetic
fields with $k \ll \mu$. The bubble diagram $ \Pi_R(t_1,t_2; k) $
takes the form, \begin{eqnarray}\label{piasi} \Pi_R(t_1,t_2; k)
&=& - 4 \int_0^{+\infty} \frac{q^4 \, dq}{(2 \, \pi)^2} \;
\int_{-1}^{+1} dx \; (1-x^2) \; \mbox{Im}\left[ G_>(t_1,t_2,q) \;
G_>(t_1,t_2, |{\vec q} +{\vec k}| ) \right]\; \theta(t_2-t_1) \cr
\cr &=&\int_0^{+\infty} \frac{q^4 \, dq}{(2 \, \pi)^2}
\;\int_{-1}^{+1} dx \; (1-x^2) \; \mbox{Im} \left[ f_q(t_1) \;
{\bar f}_q(t_2) f_{|{\vec q} +{\vec k}|}(t_1) \; {\bar f}_{|{\vec
q} +{\vec k}|}(t_2) \right]\; \theta(t_2-t_1)\; . \end{eqnarray}
where $ x \equiv {\hat q}\cdot {\hat k} $ and we set the
temperature equal to zero. For times smaller or of the order of
the nonlinear time (\ref{nonlint}) we can evaluate this bubble
diagram using eq.(\ref{alfas}) for the mode functions. This gives
for $\mu \, t \gg 1$, fixed $ t' $ and $k \ll \mu$,
$$
\mbox{Im} \left[ f_q(t) \; {\bar f}_q(t') f_{|{\vec q} +{\vec
k}|}(t) \; {\bar f}_{|{\vec q} +{\vec k}|}(t') \right]
\buildrel{\mu \, t \gg 1 , k \ll \mu}\over = e^{t (2 \, \mu -
\frac{q^2}{\mu})} \left[ 2\; |a_q^2| \;  \mbox{Im}(a_q^2) \;
e^{-t(\frac{k^2}{2\mu} + \frac{kqxt}{\mu} )} + \mbox{Im}(a_q^4) \;
e^{-t'(2 \, \mu - \frac{q^2}{\mu})} \; e^{-(t-t')(\frac{k^2}{2\mu}
+\frac{kqxt}{\mu} )} \right]\; .
$$
Calculating now the $q$ integrals in eq.(\ref{piasi}) for $ \mu \,
t \gg 1 $ yields \begin{equation}\label{piapr} \Pi_R(t,t'; k)
\buildrel{\mu \, t \gg 1 , k \ll \mu}\over= -e^2 \, N \,
\frac{(\mu^2 + M^2)}{16 \, M \, \sqrt{\mu} \, \pi^{\frac32}} \;
\frac{e^{t(2\, \mu -\frac{k^2}{4\mu})}}{t^{\frac52}} \left[1 + 2
\, e^{-t'(2\, \mu-\frac{k^2}{4\mu}) }\right]\; . \end{equation}
where we used eq.(\ref{alfas}).

Using eqs.(\ref{evatad}) and (\ref{piapr}), the photon evolution
equation (\ref{ecDC}) takes the form \begin{equation}\label{gorda}
\left(\frac{d^2}{dt^2} + k^2 + e^2 \; C^2 \; \frac{e^{2\, \mu \,
t}}{ t^{\frac32}} \right)\mathcal{D}_C(t,t'';k)-\frac{e^{t(2\, \mu
-\frac{k^2}{4\mu})}}{ t^{\frac52}}\;  e^2 \; L(t,t'') = 0\; ,
\end{equation} \noindent where \begin{eqnarray}\label{CandL} C^2 &=&
\frac{N}{64 \, \pi^{\frac32}} \; \frac{\mu^2 + M^2}{M \;
\sqrt{\mu}} \cr \cr L(t,t'') &=& \frac{N(\mu^2 + M^2)}{16 \, M \,
\sqrt{\mu} \, \pi^{\frac32}} \;\int_{t}^{t''}dt' \; \left[1 + 2 \,
e^{-t'(2\, \mu-\frac{k^2}{4\mu}) } \right] \mathcal{D}_C(t',t'';k)
\end{eqnarray} Eq.(\ref{gorda})  is valid for times larger than $ 1/\mu $
and smaller or approaching the nonlinear time $ t_{nl} $ [see
eq.(\ref{nonlint})]. It can be considered an \emph{inhomogeneous}
equation for $\mathcal{D}_C(t',t'';k)$, where the inhomogeneity is
to be determined self-consistently.

Let us first consider the homogeneous version of equation
(\ref{gorda}), \begin{equation}\label{homo}
 \left(\frac{d^2}{dt^2} + k^2 + e^2
\; C^2 \; \frac{e^{2\, \mu \, t}}{ t^{\frac32}} \right)\phi(t) =0
\end{equation} This equation has as general solution for $ \mu \, t \gg 1 $
and $k \ll \mu$ \begin{equation}\label{solhomo}
\phi(t)\buildrel{\mu \, t \gg 1 }\over= t^{\frac38}
\;e^{-\frac{\mu \, t}{2} } \left[ A \; \cos\left(\frac{e \, C}{\mu
\; t^{\frac34}} \; e^{\mu \, t} \right) + B \; \sin\left(\frac{e
\, C}{\mu \; t^{\frac34}} \; e^{\mu \, t} \right)\right]
\end{equation} where $A$ and $B$ are arbitrary constants.

In order to solve eq.(\ref{gorda}) we use the Green function of
eq.(\ref{homo})
$$
\left(\frac{d^2}{dt^2}  + k^2 + e^2 \; C^2 \; \frac{e^{2\, \mu \,
t}}{ t^{\frac32}} \right)G(t,t') = \delta(t-t')
$$
that can be written explicitly as, \begin{equation}\label{fg}
G(t,t') = -\frac{1}{e\, C} \; (t\,t')^{\frac38}
\;e^{-\frac{\mu}{2}(t+t') } \sin\left(\frac{e \, C}{\mu \;
t_<^{\frac34}} \; e^{\mu \, t_<} \right)\; \cos\left(\frac{e \,
C}{\mu \; t_>^{\frac34}} \; e^{\mu \, t_>} \right) \end{equation}
We can thus write the antisymmetric solution of eq.(\ref{gorda})
as follows,
\begin{equation}\label{solgor} \mathcal{D}_C(t,t'';k) = \frac{1}{e\, C} \;
(t\,t')^{\frac38} \;e^{-\frac{\mu}{2}(t+t') } \sin\left[ \frac{e\,
C}{\mu} \left(\frac{e^{\mu \, t}}{t^{\frac34}} - \frac{e^{\mu \,
t''}}{t''^{\frac34}} \right) \right] + e^2 \; \int dt' \left[ \;
G(t,t') \; \frac{ e^{ t'(2\, \mu - \frac{k^2}{4 \mu}) }}{
t'^{\frac52}}\; L(t',t'') - (t \leftrightarrow t'') \right]
\end{equation} The first term provides an antisymmetric solution
of eq.(\ref{homo}) which is normalized according to
eq.(\ref{vincu}). The second term comes from the inhomogeneity in
eq.(\ref{gorda}).

For $ t \sim t_{nl} $ the first term is of the order $
\lambda^{\frac14} $. Let us now compute  the second term which
will dominate in such regime. From the explicit form of
$L(t',t'')$ (\ref{CandL}) and the solution for
$\mathcal{D}_C(t,t'';k)$ given by eqn. (\ref{solgor}) it is clear
that $L(t',t'')$ is a slowly varying function of its arguments. In
particular we can set $t'=t_{nl}$ and take $L(t',t'')\sim
L(t_{nl},t'')$ outside of the integral.

Using the results given in eqs.(\ref{fg}) and (\ref{solgor}) we
find the following result, \begin{equation}\label{intasi} \int dt'
\; G(t,t') \; \frac{e^{2\, \mu \, t'}}{ t'^{\frac52}} =
\frac{e^{-\frac{k^2 \, t}{4 \, \mu}}}{ e^2 \; C^2 \; t^{1+ \frac{3
\, k^2}{16 \, \mu^2}} } \end{equation} for $ 1/\mu \ll t < t_{nl}
$. We find at $ t = t_{nl} $, \begin{equation}\label{intnl} \int
dt' \; G(t_{nl},t') \; \frac{e^{2\, \mu \, t'}}{ t'^{\frac52}} =
\frac{e^{-\frac{k^2 \, t_{nl}}{4 \, \mu}}}{ 2 \; e^2 \; C^2 \;
t_{nl}^{1+ \frac{3 \, k^2}{16 \, \mu^2}} } \end{equation} where we
have neglected rapidly oscillating phases.

We finally get collecting all factors and taking into account that
$ L(t',t'') $ varies slowly with $ t' $ compared with $ e^{2\, \mu
\, t'} $, \begin{equation}\label{Dtfin}
\mathcal{D}_C(t,t'';k)\buildrel{t_{nl}
>  t \gg 1 /\mu}\over= \frac{L(t_{nl},t'')}{C^2 \; t^{1+ \frac{3 \,
k^2}{16 \, \mu^2}} } \; e^{-\frac{k^2 \, t}{4 \, \mu}}
\end{equation} and
\begin{equation}\label{Dtnl} \mathcal{D}_C(t_{nl},t'';k)= \frac{L(t_{nl},t'')}{
2 \; C^2 \; t_{nl}^{1+ \frac{3 \, k^2}{16 \, \mu^2}} } \;
e^{-\frac{k^2 \, t_{nl}}{4 \, \mu}} \end{equation} Notice that $
e^2 $ {\bf cancels} in eqs.(\ref{Dtfin}) and (\ref{Dtnl}) since
both integrals (\ref{intasi}) and (\ref{intnl}) are proportional
to $ e^{-2} $ and the inhomogeneity in eq.(\ref{gorda}) is
proportional to $ e^{2} $. Therefore, $ \mathcal{D}_R(t,t',k) $ is
of the order $ e^0 $ for $ 1/\mu \ll t \leq t_{nl}, \; k \ll \mu
$ and fixed $ t' $.

We thus obtain for the magnetic field spectrum in the long
wavelength limit from eq.(\ref{S.fund}), \begin{eqnarray}
\label{Mnu} S_{B}(t_{nl},k)\buildrel{k\ll\mu}\over= 2 i  k^2 \;
\int_{t_0}^{t_{nl}} dt_1 \int_{t_0}^{t_{nl}}  dt_2 \;
\D_C(t_{nl},t_2; k) \; \D_C(t_{nl},t_1; k) \;
\mathbf{\Pi}_>(t_1,t_2; k) \end{eqnarray} and we used the fact
that the homogeneous contribution $S^H_B(t,k)$ in
eq.(\ref{S.fund}) can be neglected for $ t = t_{nl} $ according to
eq.(\ref{solhomo}). We can now insert eq.(\ref{Dtnl}) for $
\mathcal{D}_C(t_{nl},t'';k)$ into eq.(\ref{Mnu}). The fastly
growing factors $e^{2\, \mu \, t}$ in eq.(\ref{Pigr}) combine with
powers of $ t $ to give near $ t_{nl} \sim \frac{1}{2\mu}
\log\frac{1}{\lambda} $ a factor
$$
\frac{e^{-\frac{k^2 \,t_{nl}}{\mu}} }{\lambda^2 \; \left(\log
\frac{1}{\lambda}\right)^{\frac32+\frac{3 \, k^2}{8 \, \mu^2}}}
$$
dominating the integrals. We used here that $ L(t_{nl},t) \sim
(t_{nl}-t)^2 $ for $ t \sim t_{nl} $. This relation follows from
the fact that $\D_C(t,t'; k) \buildrel{t \to t'}\over= (t-t') $
due to eq.(\ref{vincu}) and using the relation (\ref{CandL}) for $
t \to t''$.

Performing the integrals in eq.(\ref{Mnu}) for $ t = t_{nl} $ we
obtain neglecting corrections in $1/\ln \frac{1}{\lambda} $,
\begin{equation}\label{potmag} S_{B}(t_{nl},k) \buildrel{k \ll \mu}\over =
\frac{k^2}{\mu} \; \frac{N \; e^2}{\lambda^2 \;
\left(\log\frac{1}{\lambda}\right)^{\frac32+\frac{3 \, k^2}{8 \,
\mu^2}}} \frac{1}{2^9 \; \pi^{\frac32}} \;\left(1 +
\frac{M^2}{\mu^2} \right)^2 \;   e^{-\frac{k^2 \,t_{nl}}{\mu}}
\left[ 1 + {\cal O}\left(\frac{1}{\log\frac{1}{\lambda}}\right)
\right] \end{equation} where we used eq.(\ref{CandL}).

We read in eq.(\ref{potmag}) the correlation length of the scalar
field at the nonlinear time\cite{boysinglee},
\begin{equation}\label{chi} \xi(t_{nl})=2 \;
\sqrt{\frac{t_{nl}}{\mu}} \simeq \frac{2}{\mu} \,
\sqrt{\ln\frac{1}{\lambda}} \; , \end{equation} The expression
(\ref{potmag}) for the spectrum of the magnetic field clearly
reveals that the correlation length of the magnetic field tracks
that of the scalar field during the stage of spinodal
decomposition. This result is intuitively clear since the magnetic
field is generated by the long-wavelength spinodal instabilities
corresponding to the formation of correlated domains that grow in
time during the spinodal regime as $\xi(t) \approx
\sqrt{2t/\mu}$~\cite{boysinglee}. From $S_B(t,k)$ we can extract
the magnetic energy density in long-wavelength scales equal to or
larger than $L$ through eq.(\ref{rholongwave}).

The energy density on scales $\geq L$ given by eq.
(\ref{rholongwave}) can be computed in closed form in the limits $
L\gg \xi(t_{nl})$ or $ 1/\mu \ll L\ll \xi(t_{nl})$ using
eq.(\ref{potmag}). We find \begin{equation}\label{specL}
\frac{\rho_B(L)}{\rho_\gamma} = \frac{N \; e^2}{\lambda^2 \;
\log^4\frac{1}{\lambda}} \; \frac{1} {2^{11} \; \sqrt{2} \; 5
\pi^5} \;  \left(1 + \frac{M^2}{\mu^2} \right)^2\;
\left(\frac{\mu}{T} \right)^4 \; , \quad 1/\mu \ll L\ll
\xi(t_{nl}) \; , \end{equation} and \begin{equation}\label{specL2}
\frac{\rho_B(L)}{\rho_\gamma}= \frac{N \; e^2}{\lambda^2 \;
\log^{\frac32}\frac{1}{\lambda}} \; \frac{3}{32 \; \sqrt{\pi}} \;
\left(1 + \frac{M^2}{\mu^2} \right)^2 \; \frac{1}{\mu \, L} \;
\frac{1}{\left(L \, T \right)^4} \; , \quad L\gg \xi(t_{nl})
\end{equation} for the energy on macroscopic scales . This latter
can be interpreted as energy on magnetic fields coherent on scales
of the order or bigger than $L$. We see that large scale magnetic
fields are strongly suppressed as $L^{-5}$.

\bigskip

We now compute the spectrum of the electric field from
eq.(\ref{finalEcor}). We evaluate the integrals is
eq.(\ref{finalEcor}) by the same lines as the eq.(\ref{Mnu}) using
eqs.(\ref{Dtfin}) and (\ref{Pigr}) for  $D_C(t,t'; k) $ and
$\Pi_>(t_1,t_2;k)$, respectively. We find that the end-point near
$t_{nl}$ dominates the integrals with the result \begin{equation}
S_{E_T}(0,t_{nl}) \sim \frac{N \; e^2 \; \mu}{\lambda^2} \;
\frac{1}{\sqrt{\log\frac{1}{\lambda}}} \end{equation} and
therefore,
\begin{equation}\label{specratio} \frac{S_B(k,t_{nl})}{S_{E_T}(k,t_{nl})} \sim
\frac{k^2}{\mu^2} \; \frac{1}{\log\lambda} \end{equation} Thus we
see that for long-wavelengths $k\ll \mu$ the strength of the
electric fields generated are much larger than those of the
magnetic fields. This result clearly indicates a violation of
equipartition as a consequence of the non-equilibrium generation
of electromagnetic fields. If spinodal decomposition occurs from a
vacuum state, namely the $\sigma$ field rolls from the top of the
potential and the charged fields are in a vacuum state, the
non-equilibrium processes generate mainly \emph{electric} photons.

\subsection{Spinodal decomposition in a high temperature plasma}

We now study the case of a rapid (quenched) phase transition from
an initial high temperature phase to a final low temperature
phase. Before the phase transition the system is in (local)
thermodynamic equilibrium at a temperature $T>>T_c$ with $T_c
\propto \mu/\sqrt{24\lambda}$ and the long-wavelength $k\ll T$
modes acquire a thermal mass~\cite{kapleb}
\begin{equation}\label{MT} m_T = \sqrt{\frac{\lambda}{24}}~ T .
\end{equation} The occupation number of a mode of momentum $q$ is given by
the Bose-Einstein distribution function
\begin{equation}\label{occu} n(q) =
\frac{1}{e^{\frac{\omega(q)}{T}}-1}~~;~~ \omega(q)=
\sqrt{q^2+m^2_T} \end{equation} Short wavelength modes with
momenta $q>>\lambda T$ remain in local thermodynamic equilibrium
during the (quenched) phase transition, while long-wavelength
modes undergo critical slowing down~\cite{critslowdown}, freeze
out and fall out of equilibrium. These long-wavelength modes will
undergo spinodal decomposition and their mode functions for
$\mu^{-1}< t \leq t_{nl} $ will grow as (see eq. (\ref{alfas})
\begin{equation}\label{appr.f}
 f_q(t)\simeq -i\; a_q \; e^{[\mu-q^2/(2\mu)]t},\quad
a_q \sim a_0 \sim \frac{\sqrt{m_T}}{2\mu}  \;. \end{equation}
\noindent since for $ T^2\gg T_c^2=\frac{24}\lambda \mu^2$ it
follows that $ m_T\gg \mu$.

In a high temperature plasma, charge fluctuations lead to a large
conductivity in the medium. In equilibrium the conductivity is
obtained from the imaginary part of the photon polarization and it
is dominated by particles of momenta $p\sim T$ in the loop with
exchange of photons of momenta $eT < k \ll T$~\cite{baym,yaffe}.
The Drude conductivity for an ultrarelativistic plasma at
temperature $T$ much larger than the mass of the charged fields
(in this case $m_T \sim \sqrt{\lambda}T \ll T $) is given by
\begin{equation}\label{drude} \sigma_c = \frac{e^2 \mathcal{N}(T) \; \tau}{T}
\end{equation} \noindent with $\mathcal{N}(T)\propto T^3$ is the number of
particles plus antiparticles in the plasma and $\tau$ is the
transport relaxation time. A naive estimate based on the
scattering amplitude with the exchange of photons or charged
particles with momenta $p\sim T$ would indicate that $\tau^{-1}
\sim \mathcal{N}(T)~ \alpha^2/ T^2 \sim \alpha^2 T$ (with
$\alpha^2/T^2$ the typical cross section from the exchange of a
particle with $p\sim T$) and would lead to a Drude conductivity of
the form \begin{equation}\label{appxdrud} \sigma \sim
\frac{T}{\alpha} \end{equation} However, a careful analysis
including Debye (electric) and dynamical (magnetic) screening via
Landau damping leads to the conclusion that the conductivity is
given by~\cite{baym,yaffe} \begin{equation}\label{sigmacond}
\sigma = \frac{\mathcal{C} \; N \;
T}{\alpha\ln\left[\frac{1}{\alpha{N}}\right]} \end{equation}
\noindent with $N$ the number of charged fields and
$\mathcal{C}\sim \mathcal{O}(1)$. Thus  we consider separately the
contributions to the photon polarization from loop momenta in the
two very different regimes: a) the hard momenta $p \sim T \gg \mu
$ correspond to charge fluctuations that are always in local
thermodynamic equilibrium, b) the soft momenta $p \ll \mu$ fall
out of equilibrium and undergo long-wavelength spinodal
instabilities. The contribution from hard momenta will lead to a
large \emph{equilibrium} conductivity in the medium, while the
contribution to the polarization from soft momenta will contain
all the non-equilibrium dynamics that lead to the generation of
electromagnetic field fluctuations.

As the instabilities during the phase transition develop, the
fluctuations of the charged fields will generate non-equilibrium
fluctuations in the long-wavelength components of the electric and
magnetic fields and the ensuing generation of long-wavelength
magnetic fields. However, the large conductivity of the medium
will hinder the generation of electromagnetic fluctuations, hence
the conductivity must be fully taken into account to assess the
spectrum of the magnetic and electric fields generated during the
non-equilibrium stage.

In equilibrium the long-wavelength and low frequency limit
($k,\omega \rightarrow 0$) of the spatial and temporal Fourier
transform of the transverse polarization is given by
\begin{equation}\label{equi} \Pi_T(k,\omega)= i\omega \sigma \end{equation} Thus we write
for the full transverse polarization for long-wavelength
electromagnetic fields \begin{equation}\label{pola} \Pi_T(t,t',k)=
\sigma \; \frac{d}{dt'}\delta(t-t') + \Pi_{noneq}(t,t',k)
\end{equation} \noindent with $\Pi_{noneq}(t,t',k)$ the
contribution from the spinodally unstable long-wavelength modes
given by equations (\ref{leadtad}). Our strategy is to obtain the
non-equilibrium contribution to the spectrum of electromagnetic
fields to lowest order in $\alpha$ but treating the conductivity
\emph{exactly}.

The zeroth order propagators are now obtained by considering the
contribution from the conductivity to the equations of motion, the
relevant retarded, advanced and symmetric transverse correlators
obey \begin{eqnarray} &&\left[\frac{d^2}{dt^2}+k^2+\sigma
\frac{d}{dt}
\right]\mathcal{D}^{(0)}_R(t,t',k)=\delta(t-t')~~;~~\mathcal{D}_R(t,t')=0~~
\mathrm{for}~t<t'\label{retcon}\\
&&\left[\frac{d^2}{dt^2}+k^2+\sigma \frac{d}{dt}
\right]\mathcal{D}^{(0)}_A(t,t',k)=\delta(t-t')~~;~~\mathcal{D}_A(t,t')=0~~
\mathrm{for}~t>t'\label{advcon}\\
&&\left[\frac{d^2}{dt^2}+k^2+\sigma \frac{d}{dt}
\right]\mathcal{D}^{(0)}_H(t,t',k)=0 \label{homocon}
\end{eqnarray} Since $\sigma \sim T/\alpha$ and in the initial
state $T > T_c \sim \mu/\sqrt{\lambda}$ then for time scales in
the intermediate regime $\mu^{-1}<t<t_{nl} $ it is clear that $t
\gg \sigma^{-1}$.

For $k\ll \sigma$ and $t\gg 1/\sigma$ (when we can neglect the
second order time derivatives in (\ref{retcon}-\ref{homocon})) we
find \begin{eqnarray} &&\mathcal{D}^{(0)}_R(t,t',k)=
\frac{e^{-\frac{k^2}{\sigma}(t-t')}}\sigma\;\theta(t-t') \quad ,
\quad \mathcal{D}^{(0)}_A(t,t',k)=-
\frac{e^{-\frac{k^2}{\sigma}(t-t')}}\sigma\;\theta(t'-t)
\label{advasig} \\
&&\mathcal{D}^{(0)}_H(t,t',k)=
i\;\frac{e^{-\frac{k^2}{\sigma}(t+t')}}\sigma  \quad , \quad
\mathcal{D}^{(0)}_C(t,t',k)= e^{-\frac{k^2}{\sigma}(t-t')}/\sigma
. \label{antisime} \end{eqnarray} The homogeneous solution
$F(t,t',k)$ obeys the linear integral equation
\begin{equation}\label{integeqn}
F(t,t',k)=\mathcal{D}^{(0)}_H(t-t',\vec k)+ \left\{\int^t_{t0}dt_1
\mathcal{D}^{(0)}_C(t-t_1,\vec k) \int dt_2 \left[ {\Pi}^{l}(t_2)
~\delta(t_1-t_2)+\Pi_R(t_1,t_2;\vec k) \right]F(t_2,t',k)+
t\leftrightarrow t' \right\} \end{equation} which can be solved in
perturbation theory. Writing $F(t,t',k)=F^{(0)}(t,t',k)+ \alpha \;
F^{(1)}(t,t',k)+ {\cal O}(\alpha^2)$ we find
\begin{eqnarray}\label{1stord}
&&F^{(0)}(t,t',k)=\mathcal{D}^{(0)}_H(t-t',\vec k) \label{zerothord}\\
&&F^{(1)}(t,t',k)= \int^t_{t0}dt_1 \mathcal{D}^{(0)}_C(t-t_1,\vec
k) \int dt_2 \left[ {\Pi}^{l}(t_2)
~\delta(t_1-t_2)+\Pi_R(t_1,t_2;\vec k) \right]
\mathcal{D}^{(0)}_H(t_2-t',\vec k)+ (t\leftrightarrow t')
\end{eqnarray} Obviously $F^{(0)}(t,t',k)$ gives the vacuum
contribution to the magnetic field spectrum and must be
subtracted. As discussed in the previous section both the tadpole
and the retarded (bubble) self-energy are of order $e^2/\lambda$
near the non-linear time, whereas $\Pi_>$ is of order
$e^2/\lambda^2$.

Hence one can show iterating the evolution equation in $e^2$ that
the homogeneous contribution $S^H_B(t,k)$ is subleading by one
power of $\lambda$ with respect to $S^I_B(t,k)$.

Furthermore,  since we are interested in the soft momenta regime
because only modes in the spinodal band $q^2<\mu^2\ll m_T^2$
increase, we may approximate \begin{equation}\label{T>>T.C}
1+n_q\simeq n_q\simeq \frac T{W_q}\simeq\frac
T{m_T}=\sqrt{\frac{24}{\lambda}} \simeq n_{|\vec q+k|}\simeq
1+n_{|\vec q+k|}\;. \end{equation} The long-wavelength mode
functions in the spinodally unstable band are approximately given
by (\ref{alfas}) for $0<t\leq t_{nl}$, hence in this approximation
of long-wavelength and high initial temperature  the magnetic
spectrum \begin{eqnarray}\label{S.B.longexpr}
&&S_B(t,k)=e^2\int_{\vec q}q^2(1-\cos^2\theta)\;
\left[(1+n_q)(1+n_{|\vec q+\vec k|})\left|\int_{t_0}^{t}
dt_1\;k\;\D_C(t,t_1,k ) f_q(t_1)f_{|\vec
q+\vec k|}(t_1)\right|^2+\;\right.\nonumber\\
&&\left.(1+n_q)n_{|\vec q+\vec k|}\left|\int_{t_0}^{t}
dt_1\;k\;\D_C(t,t_1,k ) f_q(t_1)f^*_{|\vec q+\vec
k|}(t_1)\right|^2+ n_q(1+n_{|\vec q+\vec k|})\left|\int_{t_0}^{t}
dt_1\;k\;\D_C(t,t_1,k ) f^*_q(t_1)f_{|\vec
q+\vec k|}(t_1)\right|^2+\;\right.\nonumber\\
&&\left.n_qn_{|\vec q+\vec k|}\left|\int_{t_0}^{t}
dt_1\;k\;\D_C(t,t_1,k ) f^*_q(t_1)f^*_{|\vec q+\vec
k|}(t_1)\right|^2\;\right]\;.\nonumber \end{eqnarray} simplifies
to
\begin{equation}\label{S.B.high.T} S_B(t,k)=\frac{96\alpha N~k^2}{\pi \lambda
~\sigma^2} \; e^{-\frac{2k^2}{\sigma}t}\int q^4 dq~ d(\cos \theta)
(1-\cos^2\theta)\left|\int_{t_0}^{t}e^{\frac{k^2}{\sigma}t_1}
f_q(t_1)f_{|\vec q+\vec k|}(t_1)dt_1\right|^2\;. \end{equation}
The integrals over momenta and angles can be done
straightforwardly in the limit $k\ll \mu$. By using the high
temperature expression for the nonlinear time \rif{nonlint} we
find that the integral is dominated by the upper limit and we
obtain
\begin{equation}\label{finSBhiT} S_B(k,t\sim t_{nl}) \sim
\frac{4\pi^{5/2}N\alpha~k^2 \xi(t_{nl})}{\lambda^2}~
e^{-k^2\xi^2(t_{nl})/4} \end{equation} \noindent with the
correlation length or domain size given by $\xi(t)=
\sqrt{\frac{2t}{\mu}}$ but now with the  nonlinear time given by
$t_{nl} = \ln(\frac{64\mu}{\lambda T})/2\mu$.

An important aspect of this result is that the factor
$e^{-2k^2t/\sigma}$ is cancelled by a similar factor in the time
integrand, this is a consequence of the fact that at long time $t
\gg 1/\mu$ the integral is dominated by the upper limit. This
cancellation in turn implies that the magnetic field spectrum
generated by the instabilities is \emph{insensitive to the
diffusion length} $\xi_{diff}(t)\sim\sqrt{t/\sigma}$.

Besides some numerical factors and the arguments in the
logarithms, the most important aspect as compared to the case of a
quench in vacuum from the previous section is the factor
$\frac{\mu^2}{\sigma^2} $. Writing $T=T_c (T/T_c)$ and with $T_c =
\mu \sqrt{24/\lambda}$ we find \begin{equation}\label{musigratio}
\frac{\mu}{\sigma}\sim \sqrt{\frac{\lambda}{24}} \frac{\alpha~ T_c
\ln(1/N\alpha)}{N~T} \ll 1 \end{equation} Therefore, the presence
of a high conductivity plasma severely hinders the generation of
magnetic fields. However, a noteworthy aspect is that up to the
nonlinear time the magnetic field is still correlated over the
size of the scalar field domains rather than the diffusion length
$\xi_{diff} \approx \sqrt{t/\sigma}$. This is an important point,
the free field power spectrum for long-wavelengths $k\ll \sigma$
in a medium with high conductivity is given by
\begin{equation}\label{FFB} S^{(0)}_B(t,k) = -i\;k^2
D^{(0)}_H(t,t,k)= \frac{k^2}{\sigma}~ e^{-\frac{2 k^2t}{\sigma}}
\end{equation} \noindent which clearly displays the diffusion length scale
$\xi_{diff}=2 \sqrt{t/\sigma}$. The diffusion length typically
determines the spatial size of the region in which magnetic fields
are correlated in the absence of non-equilibrium generation. The
ratio between the domain size $\xi(t)$ and the diffusion length
scale $\xi_{diff}(t)$ is given by
\begin{equation}\label{scalesratio} \frac{\xi(t)}{\xi_{diff}(t)}
\sim \sqrt{\frac{\sigma}{\mu}}\gg 1 \end{equation} Where we have
used the estimate (\ref{musigratio}). Thus an important conclusion
of this study is that the magnetic fields generated via spinodal
decomposition are correlated over regions comparable to the size
of scalar field domains which are \emph{much larger} than the
diffusion scale. While this study does not directly apply to the
cosmological situation, it is definitely encouraging and will be
studied in more detail elsewhere\cite{magfieldII}.

The spectrum for the  electric field  can be obtained from that of
the magnetic field by  simply replacing $k\;\D_C \to\dot \D_C$. In
the soft regime and for time scales $\frac1\sigma\ll t\ll
\frac{\sigma}{k^2}$ we have $ \dot \D_C\simeq-k^2/\sigma^2 $
whereas $k \,\D_c\simeq k/\sigma$. Therefore the electric field
spectrum is suppressed by a factor $k^2/\sigma^2$ with respect to
the magnetic field, namely \begin{equation}\label{hiTelec}
S^{\sigma}_E(t,k)=\frac{k^2}{\sigma^2} \; S^{\sigma}_B(t,k)\;.
\end{equation} Thus in a high temperature plasma with large
conductivity the non-equilibrium processes favor the generation of
magnetic photons instead of electric photons, and again
equipartition is not fulfilled.

The energy density on wavelenghts $\lambda\geq L$  can be computed
in closed form in the limits $L\gg \xi(t_{nl})$ or $L\ll
\xi(t_{nl})$. The limit $L\gg \xi(t_{nl})$ is the relevant one, if
one is interested in the generation of large scale magnetic
fields; the limit $L\ll \xi(t_{nl})$ is important in order to
estimate the power  on small scales, this is important in
cosmology since anisotropies in the cosmic microwave background
imply severe constraints on the strength of magnetic fields at
short scales.

Since we have used the expression for the mode functions in the
spinodal band, small scales  means scales much smaller than $\xi$
but still larger than $\mu^{-1}$.

We find, \begin{equation}\label{specLcond} \rho_B(L)= 512 \;
\pi^{\frac{15}{2}} \; \frac{N \; \alpha}{\lambda^2} \;
\frac{\mu^2}{\sigma^2} \left\{\begin{array}{c}
  \frac{1}{L^4} \; \frac{\xi(t_{nl})}{5L} ~~;~~ L \gg
  \xi(t_{nl}) \\
 \frac{12\sqrt{\pi}}{\xi^4(t_{nl})}~~;~~ L\ll \xi(t_{nl})
\end{array}   \right.
\end{equation} The ratio of the magnetic energy density on scales larger than
$L$ at the  nonlinear time and the magnetic energy density in the
radiation background, given by the Stefan-Boltzman law
$\rho_\gamma=\pi^2 T^4/15$ is given by
\begin{equation}\label{ratiorhos} \frac{\rho_B(L)}{\rho_\gamma} =
7680 \;  \pi^{\frac{11}{2}} \; \frac{N \; \alpha }{ \lambda^2 }\;
\frac{\mu^2}{\sigma^2} \left\{\begin{array}{c}
  \frac{1}{(LT)^4} \; \frac{\xi(t_{nl})}{5L}~~;~~ L \gg
  \xi(t_{nl}) \\
 \frac{12\sqrt{\pi}}{[\xi(t_{nl})T]^4}~~;~~ L\ll \xi(t_{nl}) \; .
\end{array}   \right.
\end{equation} While the prefactor may depend on the details of the
cosmological setting, the factors $(LT)^{-4}, (\xi T)^{-4}$ are
purely dimensional and are ultimately the determinining factors
for the strength of the generated magnetic fields on a given
scale. These factors are \emph{invariant} under the cosmological
expansion and are determined by the ratio of the scales of
interest today (galactic) to the thermal wavelength (today) of the
cosmic microwave background radiation at the Wien peak.

\vspace{1mm}

{\bf What can we learn for cosmology?:}

Although in this article we have focused  on quenched phase
transitions in Minkowski space-time, we expect that several of the
results that we have found are robust and will emerge in a full
and detailed study in cosmological space times.

\begin{itemize}
\item{The enhancement $\sim 1/\lambda^2$ is expected on physical
grounds since the current-current correlation function that
determines the photon polarization is constructed of mode
functions whose amplitude become non-perturbatively large near the
end of the transition, at the  nonlinear time scale. This is an
unavoidable consequence of the classicalization of long-wavelength
fluctuations during the phase
transition\cite{nuestros,eri97,destri}. }

\item{ A quantity of relevance in the astrophysics of magnetic
fields is the correlation length of the magnetic fields. A result
that transpires from our analysis of the vacuum case and the case
of a phase transition in a high temperature, or radiation
dominated plasma, is that during the early and intermediate stages
of the transition, the correlation length of the magnetic field
tracks the size of the correlated scalar field domains. In the
case of a highly conducting plasma we have seen that the size of
domains is \emph{much larger} than the diffusion length, hence the
result obtained here is encouraging for generating magnetic fields
on scales far larger than the diffusion length.  }

\item{The large conductivity of the medium at high temperature
hinders magnetogenesis. This result is in qualitative agreement
with those in~\cite{Giovannini:2000dj} and implies that \emph{any}
calculational framework to obtain the spectrum of magnetic fields
generated through non-equilibrium processes \emph{must} account
for the conductivity. In our study the conductivity enters in the
form $\mu^2/\sigma^2$ with $\mu$ the symmetry breaking scale,
however in cosmology we expect the Planck scale to enter as well.
}

\item{The important ratio $r(L)$ that determines the viability of
a dynamo to amplify a seed magnetic field features the factor
$(LT)^{-4}$ where $L$ is the scale of interest. This factor is
expected on dimensional grounds if the power spectrum at long
wavelengths is $\propto k^2$ as would be naively expected in a
medium of high conductivity (see eq. (\ref{FFB})). This factor is
invariant under cosmological expansion and today $LT\sim
L/\lambda_{cmb}$ with $\lambda_{cmb}$ is the wavelength at the
Wien peak of the blackbody spectrum at temperature $T$, today
$T=T_{cmb}$. For a galactic scale ($\sim 30 \mathrm{Kpc}$)  $LT
\sim 10^{24}$ and for the scale of galaxy clusters ($\sim 1
\mathrm{Mpc}$) $LT \sim 10^{26}$.  Thus the ratio $r(L)$ is
naturally extremely small. Thus in order to overcome this enormous
factor, which is obviously related to the cosmological expansion
since the time of the phase transition, either the phase
transition must occur fairly late, or the non-equilibrium
processes must last for a long time or the generation of magnetic
fields must involve a dramatic amplification factor. Of particular
importance would be a phase transition in a radiation dominated
cosmology that leads to a scaling solution for the dynamical
evolution of the charged scalar fields. In this case the
correlation function of the charged fields evolve in time in a
scaling manner which will lead to a continuous generation of
magnetic fields. This possibility is tantalizing because the
$O(N)$ linear and non-linear sigma models do lead to a scaling
solution in the large $N$ limit\cite{turok,durrer,scalingboyhec}.

We will report on our study of these possibilities in detail
elsewhere\cite{magfieldII}.  }

\end{itemize}

\section{Conclusions}

In this article we have provided a framework to study consistently
the generation of primordial magnetic fields produced by
non-equilibrium phase transitions. The main premise is that during
the process of spinodal decomposition which is the hallmark of
non-equilibrium phase transitions (without free energy barriers)
the instabilities that drive the process of phase separation and
domain formation would lead to strong charge and current
fluctuations if the scalar fields carry (hyper) charge. One of the
main results of this article is a calculational framework derived
from the underlying non-equilibrium Schwinger-Dyson equations. The
result is an {\bf exact} expression for the spectrum of magnetic
and electric fields which is valid in all generality  for scalar
and spinor charged fields and in an arbitrary cosmological
background[see eq.(\ref{S.fund})]. The main ingredient is the
transverse photon polarization and we have confirmed that this
expression has the correct equilibrium limit.

Separating the contribution from short wavelengths that are in
local thermodynamic equilibrium from the long-wavelength
fluctuations which fall out of equilibrium during the phase
transition allows us to include the dissipative effects associated
with the conductivity in a high temperature plasma.

We studied the generation of magnetic fields during quenched phase
transitions in Minkowski space time for a theory of $N$ charged
scalar fields coupled to an abelian gauge field (hypercharge) and
one neutral scalar field (the inflaton). This is a reliable
prelude to cosmology since the dynamics in
Friedmann-Robertson-Walker backgrounds is similar to that in
Minkowski space-time with a time dependent mass. Symmetry breaking
occurs along the neutral direction and gauge symmetry is not
spontaneously broken. We explicitly obtained the spectrum of
electric and magnetic fields to leading order in $\alpha$
(hypercharge) and to leading order in the large $N$ [see
eqs.(\ref{potmag})-(\ref{specratio})].

Two cases of cosmological relevance were studied, a transition in
vacuum that models an inflationary phase transition, and one from
a high temperature phase to a low temperature phase that models a
transition in a radiation dominated era. In the second case the
high temperature plasma has a large conductivity. Separating the
contribution from short wavelength fluctuations to the photon
polarization we include the effects of the conductivity, while
long wavelength fluctuations of the charged fields fall out of
equilibrium during the transition and lead to the generation of
magnetic and electric fields. We have provided explicit
expressions  for the magnetic and electric power spectra for
 long wavelengths [see
eq.(\ref{hiTelec})-(\ref{ratiorhos})].

These studies have revealed several robust features which are
expected to survive in cosmological spacetimes:

\begin{itemize}
\item{The magnetic fields are correlated over length scales
comparable to the size of the scalar field domains produced during
the spinodal stage and in a high temperature plasma,  this scale
is much larger than the magnetic diffusion length. }

\item{Equipartition between electric and magnetic fields does {\bf
not} hold out of equilibrium. For rolling down from an initial
vacuum state the ratio of electric to magnetic power behaves  as $
\frac{\mu^2}{k^2} \, \log \lambda $ for long wavelengths [see
eq.(\ref{specratio})].  Electric photons overwhelm magnetic ones
in this case. In the case of a conducting plasma, the ratio
between electric and magnetic power behaves as $k^2/\sigma^2$ for
small $k$ with $\sigma$ the DC conductivity [see
eq.(\ref{hiTelec})]} therefore magnetic photons dominate over
electric ones in this case.

\item{The conductivity severely hinders the generation of
long-wavelength magnetic fields.   }

\item{The ratio $r(L)$ between the energy density of magnetic
fields on scales larger than $L$ and the energy density of
background radiation features an ubiquitous factor $(LT)^{-4}$
which leads to large suppression factor which must be overcome in
order for primordial seed fields to be amplified by the galactic
dynamo mechanism. This point was originally made in
ref.\cite{TurnerWidrow} and is one of the formidable roadblocks
that must be overcome towards the explanation of galactic and
extragalactic magnetic fields from primordial seeds. }

\end{itemize}

We will report on our study of magnetogenesis in inflationary and
radiation dominated cosmologies in a forthcoming
article\cite{magfieldII}.

\acknowledgements The authors thank Peter Biermann,  Massimo Giovannini,
Da Shin Lee, Kin Wang Ng, Richard Holman and Matt
Martin for interesting discussions. D. B. and M. S. thank N.S.F.
for support through grants PHY-9988720 and NSF-INT-9815064.

\end{document}